\documentclass[11pt]{article}

\usepackage[final]{acl}

\usepackage{times}
\usepackage{latexsym}
\usepackage{tabularx}
\usepackage{amsmath}
\usepackage{booktabs}
\usepackage{multirow}
\usepackage{array}
\usepackage{tikz}
\usepackage{tcolorbox}  
\usepackage{enumitem}
\tcbuselibrary{breakable}
\usetikzlibrary{calc}

\usepackage[table]{xcolor}
\usepackage{colortbl}
\definecolor{aclgray}{gray}{0.92}

\definecolor{ukpbg}{RGB}{248,245,240}
\definecolor{bjbg}{RGB}{245,248,255}

\definecolor{jointblue}{RGB}{75,145,217}
\definecolor{independentorange}{RGB}{230,120,0}

\definecolor{targetpaper}{RGB}{150, 110, 180}
\definecolor{citingpaper}{RGB}{48, 67, 129}
\definecolor{aclgreen}{RGB}{200,235,200}

\usepackage{tikz}
\definecolor{ballgray}{HTML}{4A4A4A}
\newcommand{\ballnumber}[1]{\tikz[baseline=(myanchor.base)] \node[circle,fill=ballgray,inner sep=1pt] (myanchor) {\color{white}\bfseries\footnotesize #1};}

\usepackage{listings}
\usepackage{xcolor}

\lstdefinelanguage{json}{
  basicstyle=\ttfamily\footnotesize,
  numbers=left,
  numberstyle=\tiny,
  stepnumber=1,
  numbersep=6pt,
  showstringspaces=false,
  breaklines=true,
  frame=single,
  tabsize=2,
  backgroundcolor=\color{gray!5}
}

\usepackage[T1]{fontenc}

\usepackage[utf8]{inputenc}

\usepackage{microtype}

\usepackage{inconsolata}

\usepackage{graphicx}
\usepackage{soul}

\newcommand{\methodname}{Crystal}
\newcommand{\datasetname}{Crystal-Bank}
\newcommand{\datasetsize}{46,890}

\title{\methodname: Characterizing Relative Impact of Scholarly Publications}

\author{
  Hannah Collison \\
  \texttt{hgonza13@jhu.edu}
  \And
  Benjamin Van Durme \\
  \texttt{vandurme@jhu.edu}
  \\
  \\
  Johns Hopkins University
  \And
  Daniel Khashabi \\
  \texttt{danielk@cs.jhu.edu}
}

\begin{document}
\maketitle 
\begin{abstract}

Assessing a cited paper's impact is typically done by analyzing its citation context \textit{in isolation} within the citing paper. While this focuses on the most directly relevant text, it prevents \emph{relative} comparisons across all the works a paper cites. We propose {\methodname}, which instead jointly ranks \emph{all} cited papers within a citing paper using large language models (LLMs). To mitigate LLMs' positional bias, we rank each list three times in a randomized order and aggregate the impact labels through majority voting. This joint approach leverages the full citation context, rather than evaluating citations independently, to more reliably distinguish impactful references. {\methodname} outperforms a prior state-of-the-art impact classifier by $+9.5\%$ accuracy and $+8.3\%$ F1 on a dataset of human-annotated citations. {\methodname} further gains efficiency through fewer LLM calls and outperforms prior baselines using an open-weight model, enabling scalable, cost-effective citation impact analysis. In a case study of ACL Test-of-Time award-winning papers, we find that \methodname{}'s impact characterizations align closely with long-term scientific recognition. We release \datasetname{}, a 46.8k-paper dataset with rankings and impact labels, along with code.





\end{abstract}

\section{Introduction}

Funding agencies, hiring committees, and researchers routinely rely on measures of \textit{scientific impact} to make consequential decisions; yet most such measures reduce a paper's influence to noisy proxies such as citation count~\cite{garfield1972citation}.
Citation count alone is a poor measure of impact since not all citations are equally important \cite[][]{zhu2015measuring, aguiris2012scholarly}. For example, a citation providing background information contributes less to the citing paper than one that adopts the cited paper's methodology \cite{Hassan2017IdentifyingIC}. 

\begin{figure}[t]
  \centering
  \includegraphics[width=0.9\columnwidth]{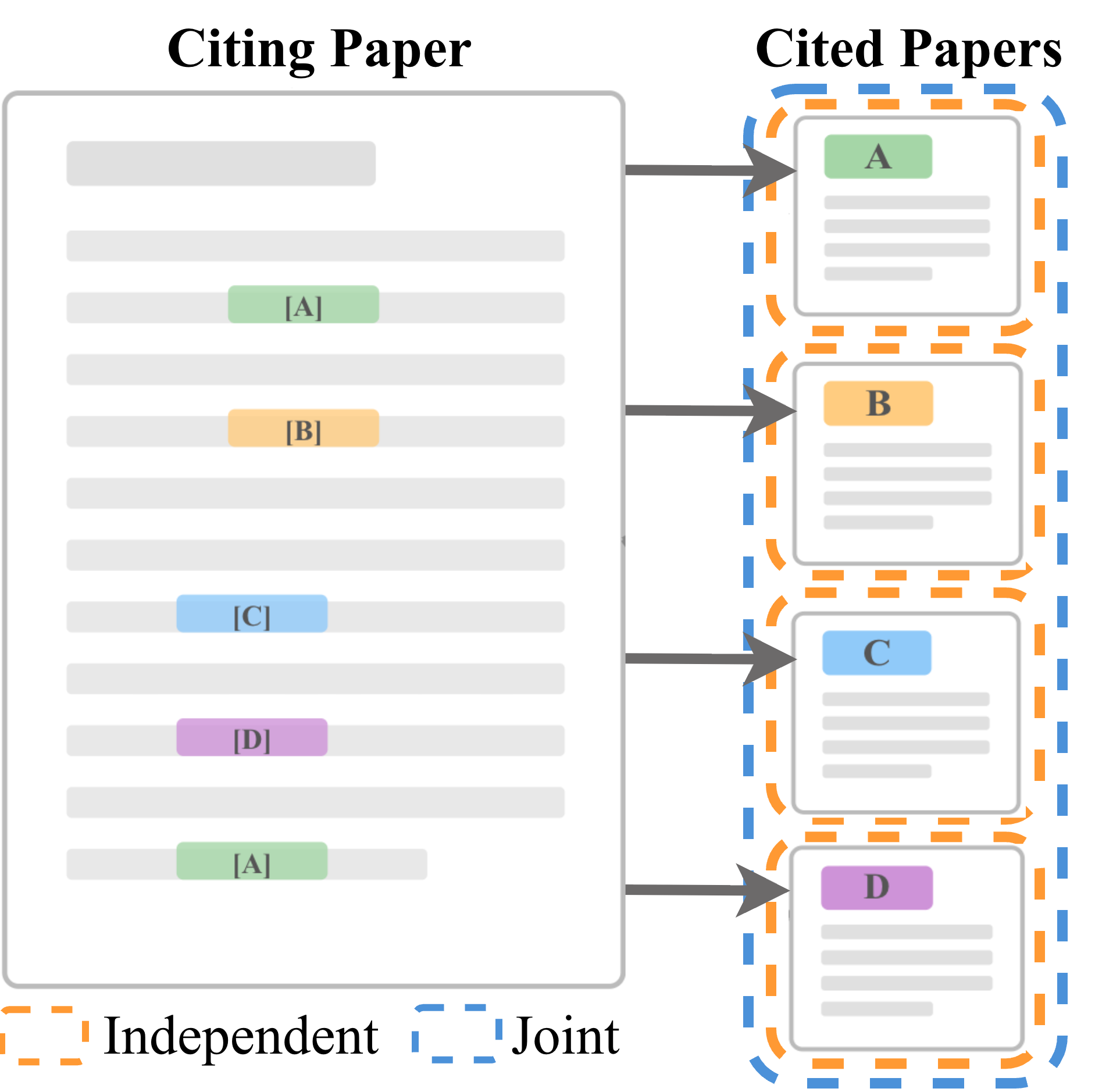}
  \caption{A citing paper $p$ (left) references papers $\{A, B, C, D\}$ (right), with each arrow denoting a citation edge $(p \rightarrow \tilde{p})$ for $\tilde{p} \in \{A, B, C, D\}$. {\methodname} considers all cited papers \textit{\textcolor{jointblue}{jointly}}
  to assess the \emph{relative} impact $\tilde{p}$ had on $p$, unlike prior work that considers only $p$ and $\tilde{p}$ in isolation, evaluating impact \textcolor{independentorange}{independently}.}
  \label{fig:1}
\end{figure}

These challenges have motivated a line of research on citation intent and impact classification \citep[e.g.,][]{valenzuela2015identifying, jurgens2018measuring, Arnaout2025IndepthRI}, which aims to assess the impact and intent behind each citation to a prior work mentioned in a given paper. As illustrated in Figure~\ref{fig:1}, these approaches evaluate each citation edge independently, assessing each citation based on its surrounding text, capturing \textit{how} and \textit{why} a paper is cited --- and to classify citations as meaningful or incidental. 
Focusing on each citation context \emph{in isolation} is a natural choice: 
it concentrates on the most directly relevant text as highlighted in Figure~\ref{fig:1}, and keeps context length short to avoid the risk of positional biases and the cost of inference.
Yet evaluating citations in isolation discards a valuable signal: the \emph{relative importance} of a reference compared to the other works cited alongside it. 

We propose {\methodname}, a method that \emph{jointly} ranks \emph{all} cited papers within a citing paper using LLMs to characterize their \emph{relative} impact (\S\ref{section:method}). 
As shown in Figure~\ref{fig:1}, jointly analyzing all citation contexts within a single citing paper introduces this comparative dimension, even though the additional context risks distracting the model. 

Surprisingly, this joint ranking approach—leveraging comparative context rather than evaluating citations in isolation—more effectively distinguishes impactful references, outperforming a prior state-of-the-art impact classifier \cite{Arnaout2025IndepthRI} by $+9.5\%$ in accuracy and $+8.3\%$ F1 on average across different LLMs on their dataset of citations with human-annotated impact labels (\S\ref{Sec:Experiments}). 
{\methodname} also offers an efficiency advantage through fewer LLM calls and outperforms prior baselines using an open-weight model, enabling scalable, cost-effective citation impact analysis (\S\ref{subsec:cost}). We conduct a case study of ACL Test-of-Time award-winning papers (\S\ref{sec:tot-case-study}), analyzing 
46.4k citations of ACL 1996 and ACL 2000 publications. We find that \methodname{}’s rankings align with long-term scientific recognition. We release \datasetname{}, a dataset with impact labels and rankings (\S\ref{crystal-bank}), along with our codebase.

\textbf{In summary,} our contributions are: \textbf{(a)} We propose {\methodname}, a method that \emph{jointly} ranks all cited papers within a citing paper using LLMs to assess their \emph{relative} impact. 
\textbf{(b)} We find that this joint ranking approach outperforms a prior state-of-the-art citation impact classifier; ablations confirm the joint formulation drives the gains.
\textbf{(c)} We show that {\methodname} is more scalable. \textbf{(d)} We observe that \methodname{}'s impact characterizations align closely with long-term scientific recognition. \textbf{(e)} We release \datasetname{}, a dataset of \datasetsize{} papers with rankings and impact labels, and our codebase.\footnote{Codebase and data will be released.}

\section{Notation and Terminology}

\begin{figure*}[t]
  \centering
  \includegraphics[width=\textwidth]{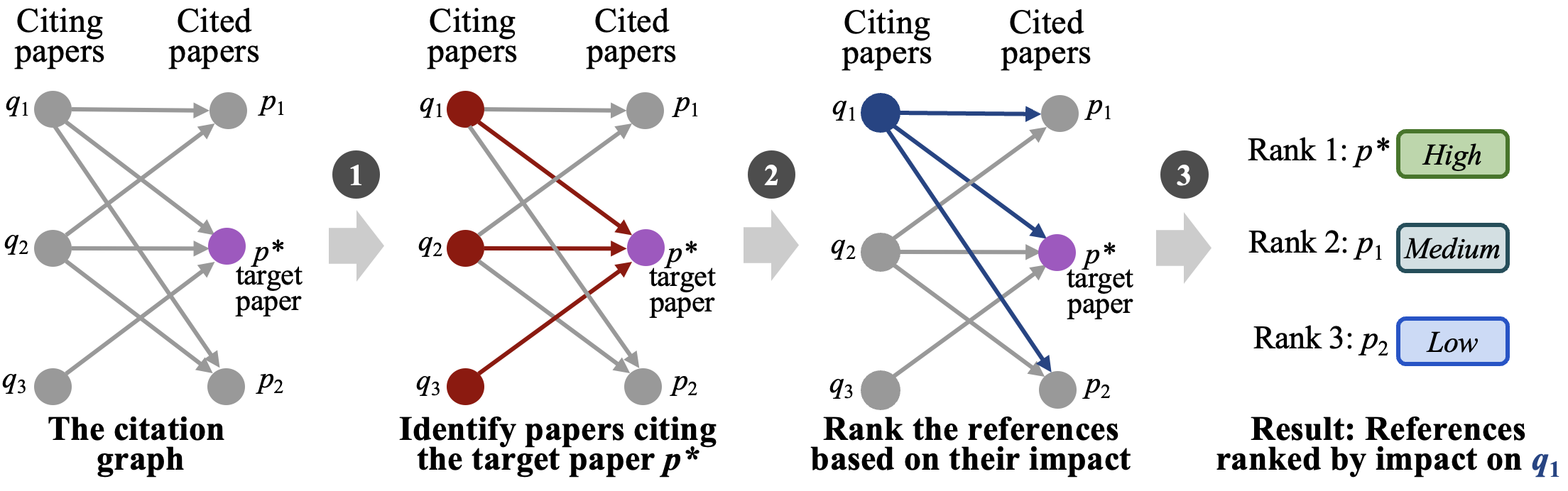}
  \caption{Overview of {\methodname}, which characterizes the relative impact of a \textcolor{targetpaper}{target paper $p^*$} on each of its citing papers, e.g.\ \textcolor{citingpaper}{$q_1$}. \protect\ballnumber{1} Identify all citing papers of \textcolor{targetpaper}{$p^*$} ($q_1$, $q_2$, $q_3$). \protect\ballnumber{2} For each citing paper, rank its references by impact (e.g., $p_1$, $p_2$, $p^*$ for $q_1$). \protect\ballnumber{3} The result is an impact-based ranking of $q_1$'s references. See \S\ref{section:method} for details.}
  \label{fig:pipeline}
\end{figure*}

\noindent\textbf{Citation graph:} We consider a universe of scholarly works
$\mathcal{P}$, where each element $p \in \mathcal{P}$ corresponds to a unique paper.
We model citations as a directed graph over $\mathcal{P}$, where a directed edge $(p \rightarrow \tilde{p})$ indicates that paper $p$ cites paper $\tilde{p}$.

Given this convention, we define two neighborhood operators.
The outgoing neighborhood of a paper $p$ is \textit{the set of papers cited by $p$}:
\begin{equation}
N_{\text{out}}(p)
\;:=\;
\left\{
\tilde{p} \in \mathcal{P}
\;:\;
(p \rightarrow \tilde{p})
\right\}.
\end{equation}
Conversely, the incoming neighborhood of a paper $p$ is \textit{the set of papers that cite $p$}:
\begin{equation}
N_{\text{in}}(p)
\;:=\;
\left\{
\tilde{p} \in \mathcal{P}
\;:\;
(\tilde{p} \rightarrow p)
\right\}.
\end{equation}

 \noindent\textbf{Citation contexts:} We assume each citation edge $(p \rightarrow \tilde{p})$ is accompanied by one or more \emph{citation contexts}, i.e., spans of text in $p$ that refer to $\tilde{p}$. We denote by $\mathrm{Ctx}(p,\tilde{p})$ the set of all such contexts:
\begin{equation}
\mathrm{Ctx}(p \rightarrow \tilde{p})
\;:=\;
\left\{
c^{(p \rightarrow  \tilde{p})}_1, \dots, c^{(p \rightarrow  \tilde{p})}_{k}
\right\},
\end{equation}
where each $c^{(p \rightarrow  \tilde{p})}_j$ is a sentence- or paragraph-level excerpt from $p$ containing an in-text citation to $\tilde{p}$.

We also define the collection of citation contexts of \emph{all} references made within a citing paper $p$:
\begin{equation*}
\mathrm{Ctx}_{\text{all}}(p)
\;:=\;
\left\{
\mathrm{Ctx}(p \rightarrow q)
\;:\;
q \in N_{\text{out}}(p)
\right\}.
\end{equation*}
Intuitively, $\mathrm{Ctx}_{\text{all}}(p)$ provides a paper-level context which we use to provide better calibrated judgments 
of impact for any particular citation $(p \rightarrow q)$.

\section{Prior Work and Broader Context}
\noindent\textbf{Citation Content Analysis:} Early studies conduct \textit{manual} analysis to assess citation quality. 
For example, \citet{moravcsik1975some} propose a four-dimensional framework for citation function and quality, which we use to define citation impact (\S\ref{llm-judge}). However, manual analysis does not scale with the rapid growth of the literature.
Natural language processing tools have thus been applied to address this challenge.
\citet{Kinney2023TheSS} release a platform that extracts citation contexts from papers.
This extraction enables downstream citation analysis tasks, such as citation classification.\\
\noindent\textbf{Citation Impact Classification:}
Several prior works, including \citet{jurgens2018measuring}, \citet{cohan2019structural}, and \citet{lauscher2022multicite}, classify \textit{citation intents} based on function (e.g., background, method, or result). In contrast, our work characterizes \emph{citation impact}, i.e., the degree to which a cited work influences the citing paper. While related, intent and impact are fundamentally distinct, and mapping between them is non-trivial (Appendix~\ref{app:intent-vs-impact}). Motivated by this distinction, prior work has proposed direct approaches to characterize citation impact. 
For example, \citet{valenzuela2015identifying} use a feature-based supervised model, and \citet{Arnaout2025IndepthRI} use an LLM judge to classify citation impact.
However, these approaches evaluate each citation \textit{independently} based on $\mathrm{Ctx}(p \rightarrow \tilde{p})$, without comparing it to other references in the same paper. 
Our approach uses $\mathrm{Ctx}_{\text{all}}(p)$, the full set of citation contexts in a paper, to calibrate each reference against others in the same citing paper.\\
\noindent\textbf{Applications of Citation Content Analysis:}
Identifying impactful citations is an important task, as it can enhance other applications, such as assessing scientific novelty \cite{shahid2025literature}, improving retrieval \cite{garikaparthi2025mir}, and tracing key contributions \cite{zhang2024pst}. It also enables more nuanced evaluation by distinguishing substantive from perfunctory citations \cite{manchanda-karypis-2021-evaluating}, and informs benchmarks for multi-dimensional impact prediction \cite{zhu2026sciimpact} and prior work forecasting \cite{ajith2026prescience}.
\section{Evaluating Relative Impact via \methodname}
\label{section:method}

We present {\methodname}, our approach for characterizing the relative impact of citations. 
As illustrated in
Figure~\ref{fig:pipeline}, the method proceeds as follows:

\noindent\textbf{(1) Downstream corpus retrieval:} Given a target paper $p^*$, we
use the Semantic Scholar API~\cite{Kinney2023TheSS} to retrieve all papers
that cite it, denoted $N_{\text{in}}(p^*)$. In Figure~\ref{fig:pipeline}, $N_{\text{in}}(p^*)=\{q_1, q_2, q_3\}$.
\noindent\textbf{(2) Citation context extraction:} For each downstream paper
$q \in N_{\text{in}}(p^*)$, we extract its full
reference list $N_{\text{out}}(q)$ and their associated citation contexts $\mathrm{Ctx}_{\text{all}}(q)$ via the same API. Figure~\ref{fig:pipeline} shows this process for $q_1$, where $N_{\text{out}}(q_1)=\{p_1, p^*, p_2\}$. The API returns the parsed sentence containing each citation; when multiple references are co-cited in a single sentence, each is assigned that full sentence as its context~\cite{Kinney2023TheSS}.

\noindent\textbf{(3) Citation-calibrated impact labeling:} 
We assign an \emph{impact label} to each citation edge $(q \rightarrow p)$ for every cited paper $p \in N_{\text{out}}(q)$.
Formally, we define an impact labeling function as:
\begin{equation*}
\label{eq:impact-labeler}
f(q \rightarrow p)
\;:=\;
f\!\big(
\mathrm{Ctx}(q \rightarrow p) | 
\mathrm{Ctx}_{\text{all}}q)
\big)
\;\in\;
\mathcal{L},
\end{equation*}
where $\mathcal{L}$ is a discrete label space.



In practice, $f$ is implemented as an LLM-based judge \cite[inter alia]{Vital2024PredictingCI, ikoma-matsubara-2023-use, lahiri2023citeprompt, koloveas2025can} and in the simplest setting, it uses binary labels $\mathcal{L}=\{0,1\}$ for \textit{impactful} and \textit{not impactful}.
However, we adopt $\mathcal{L}=\{0,1,2\}$ for \textit{low}, \textit{medium}, and \textit{high} impact, as ablations show this finer granularity improves reliability in LLM predictions by reducing high-impact false positives (\S\ref{app:two-labels}).

The key modeling choice in \eqref{eq:impact-labeler} is that the impact label for a citation $(q \rightarrow p)$ is predicted \emph{relative to all other references in $q$}.
This enables calibration against paper-specific citation conventions, such as whether $q$ tends to cite prior work superficially or relies heavily on a few core references.

In our approach, the LLM judge simultaneously ranks $q$'s references by their
impact, as shown in Figure~\ref{fig:ranking_prompt}. Prior work by \citep{wang2024large, tang2024found} demonstrates that LLMs exhibit position bias that can affect listwise ranking. We mitigate this bias with the Permutation Self-Consistency (PSC) approach proposed in \citet{tang2024found}. We randomize the order of references and perform three independent runs of the classification, each with a different randomized reference order. We then determine the impact category of the target paper $p^*$ through majority voting~\cite{byerly2025selfconsistencyfallsshortadverse}, selecting the most frequent label across the three runs. 


Additionally, we propose an alternative approach to majority voting in which we aggregate the three generated ranking files with Reciprocal Rank Fusion \cite{cormack2009reciprocal} and predict citation impact using an ordinal regression model \cite{pedregosa2015feature}. Please refer to \S\ref{Sec:AlternativeApproach} for details.

\subsection{Computational Cost}
\label{subsec:cost}

One might expect that jointly processing all citation contexts within 
a paper incurs greater computational cost than scoring each citation 
independently. We argue the opposite is true.

\noindent\textbf{Number of LLM calls:} Consider a citation graph with $n$ 
citing papers and $m$ citation edges. \methodname{} makes three LLM calls per 
citing paper, resulting in $O(n)$ calls in total. The UKP approach, 
by contrast, scores each citation edge independently, requiring $O(m)$ 
calls. Since each paper typically cites many others, $m \gg n$ in 
practice, making \methodname{} asymptotically more efficient in terms of LLM 
calls.
This advantage becomes critical at scale.
For example, OpenAlex~\cite{priem2022openalex}, comprises 248M papers and 1.9B citation edges, a scale at which \methodname{}'s $O(n)$ complexity offers a decisive computational advantage over edge-level approaches.

\noindent\textbf{Token budget:} Token budgets are comparable, as both 
approaches consume the same citation contexts. However, \methodname{} 
incurs lower prompt overhead due to its {$O(n)$} calls versus UKP's 
{$O(m)$}, meaning the prompt repeats far fewer times. In practice, this overhead is further reduced by prompt caching, making repetitions essentially free.

\section{Experimental Setup}
\label{Sec:Experiments}

\renewcommand{\arraystretch}{1.1}
\begin{table*}[t]
\centering
\small
\setlength{\tabcolsep}{8pt}
\definecolor{aclgreen}{RGB}{200,235,200}
\begin{tabular}{l l c c c c c c c}
\toprule
\textbf{Model} & \textbf{Method} & \textbf{Accuracy($\uparrow$)} & \textbf{Bal. Acc.($\uparrow$)} & \textbf{MCC($\uparrow$)} & \textbf{P($\uparrow$)} & \textbf{R($\uparrow$)} & \textbf{F1($\uparrow$)} & \textbf{F1\textsubscript{$-$}($\uparrow$)} \\
\midrule
\multirow{4}{*}{Baselines}
 & Random         & $49.9{\scriptstyle\pm1.2}$ & 50.2 & 0.00 & 33.1 & 50.9  & 40.1 & 57.0 \\
 & All Imp-Rev        & $32.9{\scriptstyle\pm1.1}$ & 50.0 & 0.00 & 32.9 & 100.0 & 49.5 & 0.0  \\
 & All Non-Imp        & $67.1{\scriptstyle\pm1.1}$ & 50.0 & 0.00 & 0.0  & 0.0   & 0.0  & 80.3 \\
 & Valenzuela     & $70.6{\scriptstyle\pm1.1}$ & 67.1 & 0.35 & 58.0 & 53.2  & 55.5 & 79.5 \\
\midrule
\multirow{3}{*}{\textsc{Gpt-5.1}}
 & UKP            & $66.7{\scriptstyle\pm1.1}$ & 65.9 & 0.30 & 49.6 & 63.5  & 55.7 & 73.4 \\
 & \methodname    & $78.6{\scriptstyle\pm1}$ & 75.8 & 0.54 & 72.2 & 63.7  & 67.7 & 85.5 \\
 & $\Delta$       & \cellcolor{aclgreen}$\mathbf{+11.9}$ & \cellcolor{aclgreen}$\mathbf{+9.9}$ & \cellcolor{aclgreen}$\mathbf{+0.24}$ & \cellcolor{aclgreen}$\mathbf{+22.6}$ & \cellcolor{aclgreen}$\mathbf{+0.2}$ & \cellcolor{aclgreen}$\mathbf{+12.0}$ & \cellcolor{aclgreen}$\mathbf{+12.1}$ \\
\midrule
\multirow{3}{*}{o4-mini}
 & UKP            & $63.0{\scriptstyle\pm1.1}$ & 65.7 & 0.29 & 46.1 & 73.6  & 56.7 & 67.7 \\
 & \methodname    & $65.4{\scriptstyle\pm1.1}$ & 74.2 & 0.53 & 76.3 & 57.1  & 65.3 & 86.0 \\
 & $\Delta$       & \cellcolor{aclgreen}$\mathbf{+2.4}$ & \cellcolor{aclgreen}$\mathbf{+8.5}$ & \cellcolor{aclgreen}$\mathbf{+0.24}$ & \cellcolor{aclgreen}$\mathbf{+30.2}$ & $-16.5$ & \cellcolor{aclgreen}$\mathbf{+8.6}$ & \cellcolor{aclgreen}$\mathbf{+18.3}$ \\
\midrule
\multirow{3}{*}{Qwen3}
 & UKP            & $60.8{\scriptstyle\pm1.2}$ & 64.9 & 0.28 & 44.5 & 76.7  & 56.3 & 64.5 \\
 & \methodname    & $75.1{\scriptstyle\pm1}$ & 71.1 & 0.46 & 70.0 & 53.3  & 60.5 & 83.9 \\
 & $\Delta$       & \cellcolor{aclgreen}$\mathbf{+14.3}$ & \cellcolor{aclgreen}$\mathbf{+6.2}$ & \cellcolor{aclgreen}$\mathbf{+0.18}$ & \cellcolor{aclgreen}$\mathbf{+25.5}$ & $-23.4$ & \cellcolor{aclgreen}$\mathbf{+4.2}$ & \cellcolor{aclgreen}$\mathbf{+19.4}$ \\
\bottomrule
\end{tabular}
\caption{Comparison of our approach \methodname{} against prior state-of-the-art method UKP~\cite{Arnaout2025IndepthRI} and four baselines: random label selection, always predicting ``impact-revealing,'' always predicting ``other,'' and the  method of \citet{valenzuela2015identifying}. Reported accuracies (in percentages) include standard error ($\pm$\,SE). Bal.\ Acc.\ = balanced accuracy; MCC = Matthews Correlation Coefficient; F1\textsubscript{$-$} = F1 on the negative (non-impact-revealing) class. \colorbox{aclgreen}{Green} cells show positive $\Delta := $ \methodname $-$ UKP; \textbf{bold} green cells show \methodname{}'s gains.  \textbf{Our approach outperforms the prior state-of-the-art method and all other baselines across all models.}}
\label{table:full_results}
\end{table*}

\noindent\textbf{Data:} Our experiments use a human-labeled dataset of citation contexts, $\mathrm{Ctx}(p,\tilde{p})$, in which each instance is annotated with a binary impact label $\mathcal{L}=\{0,1\}$ denoting whether the citation is \textit{impact-revealing} or \textit{other} \cite{Arnaout2025IndepthRI}. 

We augment the evaluation dataset with the Semantic Scholar API \cite{Kinney2023TheSS}. For every cited paper in the dataset whose title is non-empty, we obtain its Semantic Scholar identifier; for every citing paper $p$ with a valid Semantic Scholar identifier in \citet{Arnaout2025IndepthRI}, we retrieve the full reference list and all associated citation contexts $\mathrm{Ctx}_{\text{all}}(p)$. We discard citing papers whose Semantic Scholar API response contains no references, as well as duplicate entries and repeated annotations of the same citation context. 

After this filtering, the evaluation dataset \cite{Arnaout2025IndepthRI} comprises 442 citing papers, 1{,}338 cited papers, and 1{,}783 citation contexts, each annotated by human labelers as either \textit{impact-revealing} or \textit{other}. There are 1{,}196 instances (67.1\%) of the \textit{other} class and 587 (32.9\%) of the \textit{impact-revealing} class. Refer to \S \ref{limitations} for details on the type of papers in the data.



\noindent\textbf{Baselines:} We compare against five baselines. We include a random classifier, a model always predicting ``impact-revealing,'' and one always predicting ``other.'' We further compare against the supervised classifier of \citet{valenzuela2015identifying} and UKP~\citep{Arnaout2025IndepthRI}, the prior state-of-the-art LLM-based method for identifying impactful citations. We replicate UKP~\citep{Arnaout2025IndepthRI} directly following the authors' method, and obtain \citet{valenzuela2015identifying} impact predictions from the Semantic Scholar API's \texttt{isInfluential} field \cite{Kinney2023TheSS}, which Semantic Scholar documents as being generated by this classifier.\footnote{\url{https://www.semanticscholar.org/faq/influential-citations}}

We do not compare to citation \textit{intent} classifiers \citep{cohan2019structural,jurgens2018measuring,lauscher2022multicite}, as mapping \textit{intent} labels (e.g., background, method, result) to \textit{impact} labels inevitably introduces assumptions that may unfairly favor or penalize particular methods. A more detailed discussion of this distinction is provided in \S\ref{app:intent-vs-impact}.

\noindent\textbf{LLM-based Judge:} 
\label{llm-judge}
We use an LLM-based judge to assign the impact labels to citation contexts $\mathrm{Ctx}(p,\tilde{p})$ and rank the cited papers by impact for a given citing paper $p$. 
Our definition of impact is provided in Figure~\ref{fig:ranking_prompt}, which also details all inputs to the model. We validate the LLM prompt in a pilot study \S\ref{sec:pilot_study}, where we asked annotators to rank their own paper's references by impact; rankings produced by our prompt closely align with these author judgments, achieving Spearman $\rho > 0.7$ across all annotators.

We compare several LLMs of different families and sizes for this task. Specifically, we use \textsc{Gpt-5.1} and o4-mini as our closed-source models, and Qwen3-30B-A3B-Instruct-2507-FP8 \cite{qwen3technicalreport} as our open-weight model. We adopt the recommended temperature and top-$p$ settings for each model (0.7 and 0.8 for Qwen; defaults for \textsc{Gpt-5.1} and o4-mini). These models all support large context windows of at least $200K$ tokens.



\noindent\textbf{Evaluation Measures:}
\label{EvaluationMeasures}
We evaluate our model on the human-annotated dataset released by \citet{Arnaout2025IndepthRI}, which labels citations as \textit{impact-revealing} or \textit{other}. To enable comparison across baselines, we map each system’s label set to these two categories at evaluation time; 
inference is performed in each system’s original label space, with mapping applied only post hoc (details are in \S\ref{app:label-mapping}).

We assess performance using standard classification metrics. Specifically, we compute accuracy as the fraction of citations whose predicted labels match the ground truth, and report precision, recall, and F1 score for the \textit{impact-revealing} class in Table~\ref{table:full_results}. 
We additionally report the following balance aware metrics: balanced accuracy—the average of per-class recalls, ensuring equal contribution from each class—and Matthews Correlation Coefficient (MCC), which accounts for all four quadrants of the confusion matrix and yields 0.0 for trivial single-class predictors. For completeness, we also report the F1\textsubscript{$-$} score for the \textit{other} class in Table~\ref{table:full_results}.
\paragraph{Results:} As shown in Table~\ref{table:full_results}, \methodname{} consistently outperforms prior state-of-the-art impact classifier, which we refer to as UKP \cite{Arnaout2025IndepthRI}, improving accuracy by $+9.5\%$ and F1 by $+8.3\%$ on average across all three models. \methodname{} achieves MCC scores of 0.46--0.54 and balanced accuracy of 71.1\%--75.8\%, compared to UKP's MCC of 0.28--0.30 and balanced accuracy of 64.9\%--65.9\%, \textbf{indicating far stronger discrimination across both classes.} \methodname{} consistently achieves higher precision 70.0\%--76.3\% precision versus UKP's 44.5\%--49.6\%, meaning \textbf{our predictions for impactful citations are considerably more reliable.} Importantly, Table~\ref{table:full_results} reports the standard error (SE) of accuracy, which remains low $({\approx}\,{\pm}1.0\%\text{--}1.2\%$) across all settings, indicating that \textbf{the estimates are stable.} Moreover, the observed improvements over prior work are consistently larger than the corresponding SE (e.g. $+14.3\%$), suggesting that the gains are robust. 

Compared to the method of \citet{valenzuela2015identifying}, \methodname{} achieves higher F1 across all three models ($+5.0\%$--$+12.2\%$), as well as higher MCC ($+0.11$--$+0.19$) and balanced accuracy ($+4.0\%$--$+8.7\%$), demonstrating stronger and more consistent discrimination.

We also find that Qwen3-30B is a strong alternative to \textsc{Gpt-5.1}, achieving a comparab performance at lower cost. As shown in Figure~\ref{fig:confusion_matrices}, o4-mini and Qwen3-30B achieve higher recall on UKP mainly due to a tendency to over-predict the impact-revealing class. We additionally observe that models sometimes omit references from their rankings when papers have long reference lists. We treat such cases as misclassifications in Table~\ref{table:full_results}, and even under this strict evaluation, \methodname{} outperforms all baselines. As described in \S\ref{Sec:Missing References}, this limitation is expected to diminish as models' long-context capabilities improve. Overall,  \textbf{\methodname{} outperforms prior state-of-the-art method and all baselines }with substantial gains across all settings.

\paragraph{Qualitative Error Analysis:}
\label{sec:error-analysis}
A domain expert (author) manually analyzed the 45 false positives (FP) and 166 false negatives (FN) that all three models (\textsc{Qwen3-30B}, \textsc{Gpt-5.1}, and \textsc{o4-mini}) misclassified in \methodname{}, identifying commonalities in these errors. FP contexts tend to describe the cited work in detail and credit its authors as the source of a specific method or finding, often through verbs such as \textit{proposed} or \textit{introduced} or through comparative phrasing such as \textit{similar to} or \textit{variant of}, and they are substantially longer than true negatives (median 954 vs.\ 238 characters). FN contexts show two recurring patterns. 
The first is contexts that are substantially shorter than true positives (median 280 vs.\ 1{,}110 characters), offering little surrounding text for the model to draw on. The second is contexts that group the cited work with several other references in the same citation. \textbf{\methodname{} may associate the citation with its neighbors, diluting and downgrading its impact label.}
Length statistics and examples for both error categories (FP and FN) are provided in \S\ref{app:error-analysis}. 

\section{Ablation Studies}
We perform three ablation studies to isolate the contributions of \methodname{}'s key design choices.

\paragraph{1) Three vs.\ Two Impact Labels}
\label{app:two-labels}
We ablate \methodname{}'s three-class \textit{High}/\textit{Medium}/\textit{Low} formulation against the binary \textit{impact-revealing} vs.\ \textit{other} convention of \citet{Arnaout2025IndepthRI}, using the same Qwen3-30B model and experimental setup as \S\ref{Sec:Experiments}. This isolates whether performance gains stem from label granularity or from \methodname{}'s joint ranking formulation. Figure~\ref{fig:prompt_abl_two_labels} shows the prompt used.

As shown in Table~\ref{table:abl_two_labels}, \methodname{} outperforms UKP even under the binary setting ($+8.5\%$ accuracy, $+4.0\%$ balanced accuracy, $+0.09$ MCC), confirming that \textit{the joint ranking formulation, not label count, is the primary driver of improvement.} Moving to three labels yields further gains ($+12.6\%$ in precision and $+2.0\%$ F1). Analysis of the confusion matrices (Figure~\ref{fig:cm_abl1}) indicates that the \textit{Medium} bin absorbs borderline cases that the binary model incorrectly assigns to \textit{impact-revealing}, reducing false positives and increasing precision from $57.1\%$ to $70\%$.
Thus, \textbf{finer label granularity gives the model a more calibrated decision space}, enabling it to reliably identify truly high-impact citations.

\paragraph{2) Independent Variant vs.\ Joint Ranking:}
\label{app:independent}
We ablate \methodname{}'s joint ranking formulation against an independent-input variant, where each citation context $\text{Ctx}_p$ is assessed in isolation. To match \methodname{}'s majority voting procedure, we run the independent variant three times per citation and take the majority label. Impact definitions, shown in Figure~\ref{fig:prompt_abl_independent}, are identical across both settings. 

As shown in Table~\ref{table:abl_independent}, the ablation reveals a clear trade-off between recall and calibration. Using independent input leads to a strong bias toward predicting the \textit{High} class, yielding very high recall ($91.7\%$) but poor precision ($46.8\%$) and lower overall agreement (MCC $0.40$). This is reflected in the confusion matrix (Figure~\ref{fig:cm_abl2}), where both \textit{Impact-Revealing} and \textit{Other} instances are frequently assigned to \textit{High}. In contrast, joint input substantially improves class discrimination, increasing precision ($+23.2\%$), accuracy ($+12.1\%$), and MCC ($+0.06$), while producing a more balanced prediction distribution; particularly by correctly assigning most \textit{Other} cases to \textit{Low}. Although recall decreases ($-38.4\%$), the joint formulation yields better-calibrated and more reliable predictions overall, suggesting that \textbf{shared context is critical for mitigating the overprediction bias observed in the independent setting}.

\paragraph{3) Permutation-based majority voting versus a single-pass ranking:}
\label{app:abl-mv}

We ablate \methodname{}'s permutation-based majority voting procedure vs a single-pass ranking variant that uses a different randomized reference ordering per pass. This isolates the contribution of voting from the underlying model quality, and tests whether majority voting over permuted orderings robustly mitigates positional bias.

As shown in Table~\ref{table:abl3_single_run}, individual runs produce noticeably different results across orderings for all three models, confirming that positional bias is a real source of instability. Majority voting consistently matches or exceeds the best individual run across nearly all metrics. The benefit is most pronounced for o4-mini (+$11.9\%$ accuracy, $+2.8\%$ F1) and Qwen3 ($+8.5\%$ accuracy, $+1.8\%$ F1), both of which are more sensitive to reference ordering in the single-pass setting. For \textsc{Gpt-5.1}, where individual runs are already tightly clustered, voting still yields modest gains ($+1.5\%$ accuracy, $+0.2\%$ F1) without hurting any metric. Thus, \textbf{permutation-based majority voting helps most when the underlying model is less consistent, and does not hurt when it is already stable}---making it a robust default across model families.

\paragraph{To summarize,} our ablations explain \methodname{}'s gains: \textbf{(1)} the \textit{joint} ranking formulation is the main driver of improvement
\textbf{(2)} a three-class formulation further improves precision via better calibration; and \textbf{(3)} permutation-based majority voting mitigates positional bias, especially for weaker models. 

\section{Case Study: ACL Test-of-Time Awards}
\label{sec:tot-case-study}

We study whether \methodname{}'s high-impact impact citations are correlated to long-term scientific recognition, as measured by the ACL Test-of-Time (ToT) Paper Award, which annually recognizes up to four NLP and computational linguistics papers (two published 25 years prior and two published 10 years prior) for their enduring impact on the field.\footnote{\url{https://www.aclweb.org/portal/content/announcement-2025-acl-test-time-paper-award}}

\paragraph{Data:}
\label{ssec:tot-data}

\begin{table}[h]
\centering
\footnotesize
\begin{tabular*}{\columnwidth}{@{\extracolsep{\fill}}lrrrr}
\toprule
\textbf{Dataset} & \#Papers & \shortstack{Actual \\ Citing} & \shortstack{S2 \\ API} & \shortstack{API + \\ Scrape} \\
\midrule
ACL 1996 & 577  & 29,308 & 4,216  & 18,451 \\
ACL 2000 & 1,184 & 40,681 & 6,639  & 27,997 \\
\bottomrule
\end{tabular*}
\caption{Coverage statistics for ACL 1996 and 2000 papers and their citing papers. ``\#Papers'' denotes the number of papers published at ACL in each year. ``Actual Citing'' is the total number of papers citing these publications. ``S2 API'' is the subset of citing papers with non-empty reference data returned by the Semantic Scholar API. ``API + Scrape'' includes additional citing papers recovered via website scraping, which substantially improves coverage of reference information.}
\label{tab:coverage-stats}
\end{table}

We begin by selecting three ACL papers that received the Test-of-Time (ToT) Award: \citet{berger1996maximum}, \citet{carletta1996assessing}, and \citet{gildea2002automatic}. We choose these papers because they were recognized 25 years after publication and were honored in different announcement years. \citet{berger1996maximum} and \citet{carletta1996assessing} were the first recipients in 2021, while \citet{gildea2002automatic} received the award in 2025.

We follow the approach in \S\ref{section:method} to construct the citation graph. Using the ACL-OCL dataset \cite{rohatgi2023acl}, we collect all ACL papers published in 1996 and 2000. We then identify all papers that cite them using the Semantic Scholar API \cite{Kinney2023TheSS} and retrieve their citation contexts. At scale, about 85\% of API responses lack reference details due to licensing restrictions (e.g., “Notice: The following paper fields have been elided by the publisher”). Although these details are often unavailable through the API, they are typically visible on the Semantic Scholar website. We thus supplement the API data by scraping the site to recover missing reference data and citation contexts. Table~\ref{tab:coverage-stats} displays the coverage statistics of the citing papers considered in the analysis for each year.

\paragraph{Experiment:}
\label{ssec:tot-experiment}

As described in \S\ref{section:method}, we apply \methodname{} to characterize the impact of all references in papers that cite publications from ACL 1996 and ACL 2000. Concretely, we run \methodname{} on 18,451 papers that cite ACL 1996 publications and on 27,997 papers that cite ACL 2000 publications, as summarized in Table~\ref{tab:coverage-stats}. 

We use the same Qwen3-30B model described in \S\ref{Sec:Experiments} and the same prompt \S\ref{fig:ranking_prompt} when running \methodname{}. This ensures that the citation context extraction and downstream scoring are consistent with the setup used in the main experiment \S\ref{Sec:Experiments}.

\paragraph{Results:}
\label{ssec:tot-results}

\begin{figure*}[t]
  \centering
  \includegraphics[width=\textwidth]{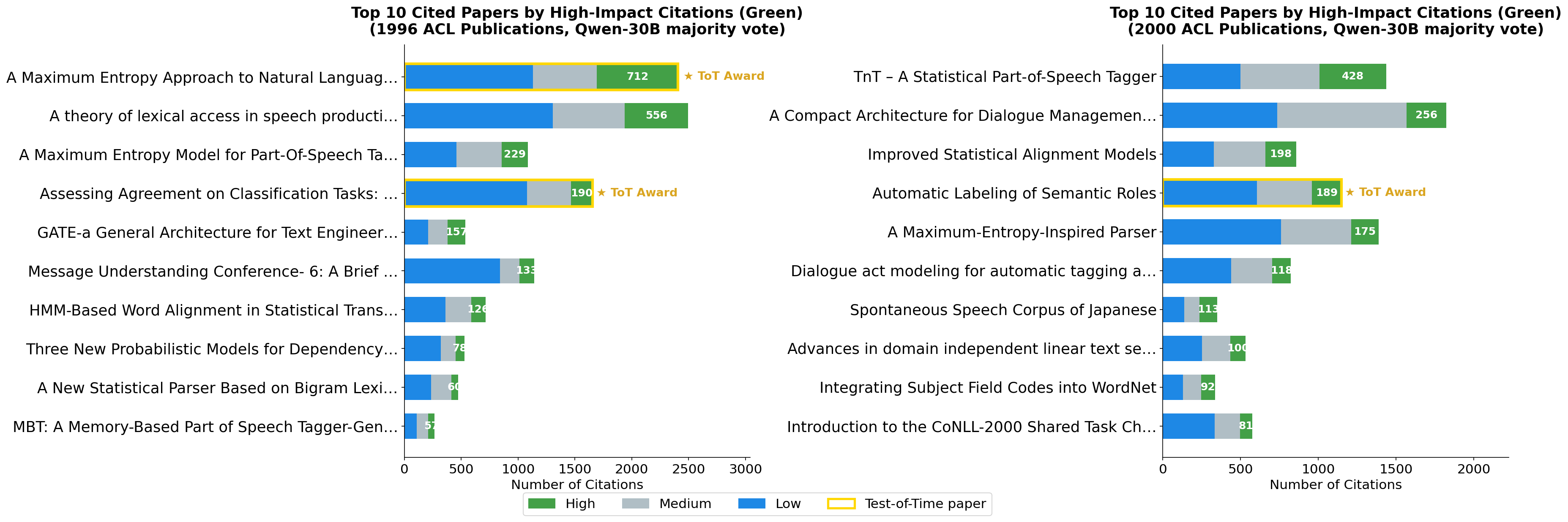}
  \caption{Impact distribution of ACL 1996 (left) and ACL 2000 (right) publications, ranked by the number of \textcolor[HTML]{43A047}{high}-impact citations as predicted by \methodname{}, showing the top 10 in each year with the top-ranked paper at the top of each plot. Each bar shows the breakdown of \textcolor[HTML]{43A047}{high}, \textcolor[HTML]{546E7A}{medium}, and \textcolor[HTML]{1E88E5}{low} impact citations per paper, with the high-impact count annotated. Papers awarded the ACL Test-of-Time (ToT) Award are highlighted with a yellow border: \citet{berger1996maximum} and \citet{gildea2002automatic}. In both years, ToT-winning papers rank among the top in \textcolor[HTML]{43A047}{high}--impact citations, suggesting that \methodname{}'s impact characterization aligns with long-term community recognition of foundational contributions.}
  \label{fig:ACL_results}
\end{figure*}

We aggregate the impact labels produced by \methodname{} for each paper published in ACL 1996 and ACL 2000. Specifically, for each paper in these two sets, we consider all papers that cite it and collect the impact labels assigned by \methodname{} to those citations. We then summarize these labels by counting, for each paper, how many of its citations are classified as high-, medium-, or low-impact. As shown in Table~\ref{tab:impact-distribution}, the label distribution is highly consistent across both years. Approximately 50--51\% of citations are labeled as low impact, 30--34\% as medium impact, and 17--19\% as high impact. This indicates that \textbf{\methodname{} assigns the high-impact label selectively}, reserving it for a relatively small fraction of citations. 

\noindent\textbf{High-Impact Citation Counts Analysis:}
We rank ACL 1996 and ACL 2000 publications by the number of high-impact citations they receive. We then examine the top 10 papers for each year. As shown in Figure~\ref{fig:ACL_results}, the ToT awarded paper \citet{berger1996maximum} ranks first with 712 high-impact citations, making it the top-ranked paper under \methodname{} in ACL 1996. Comparing this ranking to one based on raw citation counts leads to a different conclusion about which papers are most impactful. The second-ranked paper, \citet{levelt1999theory}, has a higher total number of citations but fewer high-impact citations. This difference shows that citation count alone is a weak proxy for impact, as it does not distinguish between citations that reflect substantive use and those that are merely incidental. However, \textbf{high-impact citation counts are not sufficient on their own to explain long-term recognition.} This becomes clear when comparing against ToT awards. While \citet{berger1996maximum} is both top-ranked and a ToT recipient, this alignment does not generalize. For example, \citet{carletta1996assessing} ranks fourth with 190 high-impact citations yet received the ToT award, whereas \citet{ratnaparkhi1996maximum} has more high-impact citations (229) but was not selected. A similar pattern appears in ACL 2000, where \citet{gildea2002automatic} ranks fourth despite being the ToT recipient. These cases indicate that even impact-aware aggregation does not fully capture the factors behind long-term recognition.

\noindent\textbf{Temporal Dynamics of Citation Impact:}
To better understand these differences, we examine the temporal distribution of citation impact. Figures~\ref{fig:temporal-1996} and \ref{fig:temporal-2000} show how high-, medium-, and low-impact citations are distributed year by year for the top four papers in ACL 1996 and ACL 2000, respectively. The top four papers in each year are selected by ranking all papers according to their number of high-impact citations and then taking the highest four; the ToT papers are included in this subset for both conferences. \citet{berger1996maximum} has the largest volume of high-impact citations among the ACL 1996 papers, \textit{sustained} across more than two decades. \citet{carletta1996assessing} shows a stable but smaller number of high-impact citations, with most of its citation volume concentrated in the low-impact category. For ACL 2000, \citet{gildea2002automatic} maintains a stable level of high-impact citations through 2025, despite not ranking first by aggregate counts. In contrast, the \textbf{non-ToT papers} in the top four for ACL 2000 \textbf{show high-impact citations that peak earlier and then decline more steeply.}

\textbf{In summary,} these observations highlight three distinct points. (\textbf{1}) Citation counts do not capture \textit{how} a paper is cited. (\textbf{2}) Impact characterization addresses this limitation but ignores \textit{when} that influence occurs. (\textbf{3}) Temporal patterns reveal whether impact is sustained, concentrated, or diffuse over time, which helps explain differences in long-term recognition.

Overall, \methodname{} offers a more informative view of scholarly influence than raw citation counts, though it does not fully predict ToT selections. Nevertheless, \textbf{\methodname{}'s impact characterization aligns with long-term community recognition of foundational contributions}, as all three ToT-awarded papers in our analysis rank among the top four in their respective ACL conference years by high-impact citations.

\section{Conclusion}
\label{section:results}
As shown in Table~\ref{table:full_results}, \methodname{} consistently outperforms prior state-of-the-art impact classifiers, improving accuracy by $+9.5\%$ and F1 by $+8.3\%$ on average across all three models. We also find that Qwen3-30B is a strong alternative to \textsc{Gpt-5.1}, achieving a strong performance at lower cost. 
Beyond benchmark performance, our case study (\S\ref{sec:tot-case-study}) shows that high-impact citations identified by \methodname{} align with long-term scientific recognition, as signaled by the ACL Test-of-Time Paper Award.

\textbf{In conclusion}, these results show that \textit{jointly} modeling all references within a citing paper provides richer signal than independent citation classification, and \methodname{} offers a scalable and cost-effective solution for citation impact analysis.

\section*{Limitations}
\label{limitations}
This work analyzes citing papers from psychology, medicine, and computer science. While this captures variations in the number of references in papers and citing practices across different fields, it does not comprehensively represent all disciplines. Additionally, the dataset is limited to English-language scientific papers. 

Despite these limitations, \methodname{} is broadly applicable to scholarly work containing citations. It is most naturally suited for research and position papers, where citations vary meaningfully in how central they are to the paper's argument, making both ranking and characterizing their impact informative. For survey papers, a full ranking of references may be less informative, as surveys aim for comprehensive coverage rather than building on a few key works. Nevertheless, \methodname{}'s impact categorization (high, medium, low) remains valuable for distinguishing landmark works from those included primarily for completeness.

In practice, as the list of references grows, models struggle to rank the complete list despite having sufficiently large context windows. The number of unranked references differs across models, with \textsc{Gpt-5.1} showing the best performance, suggesting this limitation will diminish as models improve. Further analysis can be found in \S\ref{Sec:Missing References}.

Also, if the Semantic Scholar API \cite{Kinney2023TheSS} fails to retrieve information for a given paper, 
the 
next steps of our pipeline cannot be completed.

Finally, an open problem is that authors sometimes cite work differently from what they perceive as impactful, as revealed by interviews in our pilot study. More details can be found in \S\ref{sec:pilot_study}.

\section*{Ethics Statement}
The dataset used in {\methodname} consists of publicly available scholarly data obtained through the Semantic Scholar API \cite{Kinney2023TheSS}, used in accordance with its terms of service. This data contains author names as part of the standard bibliographic record, but no sensitive personal information. Thus, no further anonymization was applied.

{\methodname} is intended for research purposes only. Potential future applications include generating impact reports for authors to review how their work is cited over time. We note, however, that a potential risk of characterizing impact from citation contexts is that it could incentivize authors to strategically frame their citations to influence impact scores.

\section*{Acknowledgments}
This work is supported by the Office of Naval Research (N00014-24-1-2089). We thank the JHU CLSP community for productive discussions, Zhengping Jiang for sharing an annotation UI template, and Jack Collison for his helpful feedback on this work.

\bibliography{ref_custom}

\appendix
\label{sec:appendix}

\section{Citation Intent vs. Citation Impact}
\label{app:intent-vs-impact}

The notions of citation intent and impact are fundamentally distinct, and mapping intent labels to impact labels is inherently non-trivial.
A direct comparison between intent-classification methods and impact-classification methods requires manually defining a mapping from intent categories to impact categories. However, such mappings inevitably introduce assumptions that may unfairly favor or penalize particular methods. For example, one may map \textit{background} citations to non-impactful and \textit{uses}/\textit{extends} citations to impactful, following prior work \cite{Arnaout2025IndepthRI}. However, this simplification can lead to unstable or misleading labels. To illustrate, consider the paper by \citet{yang2017breaking} in the MultiCite \cite{lauscher2022multicite} test set.\footnote{\url{https://github.com/allenai/multicite/blob/master/data/classification_gold_context/test.json}} Its annotations are evenly split across the categories \textit{background}, \textit{uses}, and \textit{extends} (5 annotations each). Under a majority-vote mapping scheme with random tie-breaking, the resulting impact label has 33\% chance of being mislabeled as non-impactful.

More importantly, citation intent does not uniquely determine citation impact. A citation unanimously labeled as \textit{background} may still be highly impactful if it introduces a foundational concept on which the citing work critically depends. For these reasons, we do not include intent-based baselines \citep{cohan2019structural,jurgens2018measuring,lauscher2022multicite} in our main comparisons and instead focus on methods explicitly designed to assess citation impact.

\paragraph{Evaluation Label Mappings:}
\label{app:label-mapping}

All systems are evaluated against the ground-truth categories of
\textit{impact-revealing} and \textit{other} from \citet{Arnaout2025IndepthRI}.
Table~\ref{tab:label-mapping} shows how each system's predicted labels are mapped
to these categories for evaluation purposes only; each system predicts
over its own native label set during inference. In \methodname{}, we map \textit{High} to
\textit{impact-revealing}, and both \textit{Medium} and \textit{Low} to
\textit{other}. This mapping reflects the definition of impact in \citet{Arnaout2025IndepthRI}, where
\textit{impact-revealing} citations explicitly emphasize the significance or
influence of prior work. While \textit{Medium} citations may indicate relevance
or usefulness, they do not consistently foreground such impact, and are therefore
grouped with \textit{other} for evaluation. Notably, we retain the three-label design (\textit{High}, \textit{Medium}, \textit{Low}), as ablations show that this finer granularity yields a more calibrated decision space, enabling more reliable identification of truly high-impact citations (\S\ref{app:two-labels}).


\begin{table}[h]
\centering
\footnotesize
\begin{tabular}{p{1.8cm}ll}
\toprule
\textbf{System} & \textbf{Predicted} & \textbf{Ground-Truth} \\
\midrule
\citeauthor{Arnaout2025IndepthRI} & Impact-revealing & Impact-revealing \\
                                  & Other            & Other \\
\addlinespace
\multirow{2}{1.8cm}{\citeauthor{valenzuela2015identifying}} & Meaningful     & Impact-revealing \\
                                                            & Non-meaningful & Other \\
\addlinespace
\methodname~(ours) & High   & Impact-revealing \\
                   & Medium & Other \\
                   & Low    & Other \\
\bottomrule
\end{tabular}
\caption{Evaluation-time label mappings to ground-truth categories
from \citet{Arnaout2025IndepthRI}, applied only at evaluation time.}
\label{tab:label-mapping}
\end{table}

\paragraph{Sample Outputs:} 
Figure~\ref{fig:ranking_prompt} shows our definition of citation impact and the instruction used to rank a citing paper's references, with a sample output in Figure~\ref{fig:sample_output_prompt}. Figure~\ref{fig:sample_output_ord_reg} displays the aggregated rankings from three independent runs using \textsc{Gpt}-5.1. We use these aggregated ranking files to predict impact labels via an ordinal regression model (\S\ref{Sec:AlternativeApproach}).

\begin{figure*}[p]
\centering
\begin{tcolorbox}[
    colback=gray!5,
    colframe=gray!50,
    boxrule=0.5pt,
    arc=4pt,
    width=\textwidth
]
\small

The task is to rank all of the references $R$ (where $R$ is a list of papers $r_1, r_2, \ldots$) of a given paper $P$ based on how impactful and influential each $r_n$ was on $P$.

\smallskip
\textbf{The paper $P$ you will be analyzing is:}
\begin{itemize}[leftmargin=1.5em, itemsep=0pt, topsep=1pt, parsep=0pt]
    \item Title: \texttt{[main\_paper\_title]}
    \item Abstract: \texttt{[main\_paper\_abstract]}
\end{itemize}

\smallskip
\textbf{The list of references is given below} (each reference includes paper ID, title, and context of citation): \texttt{[references\_text]}

\smallskip
\textbf{Impact categories:}

\smallskip
\textbf{1. High-impact citations:} These are the papers without which your own work would not have been possible. They supply essential conceptual, methodological, or operational ingredients.
\begin{itemize}[leftmargin=1.5em, itemsep=0pt, topsep=1pt, parsep=0pt]
    \item \textit{Conceptual or operational indispensability:} The reference provides a unique conceptual insight, methodological innovation, dataset, or technique that is directly instrumental to your paper. Examples: a specific algorithm your method extends; a benchmark or dataset your study critically depends on; a theoretical formulation your contribution builds on.
    \item \textit{Organic necessity:} The reference is uniquely and genuinely required for a reader to understand how your paper works or how its core logic unfolds. Without this citation, the intellectual lineage of your method would be opaque or incomplete.
    \item \textit{Typical quantity:} 1--5 papers (or even 1).
\end{itemize}

\smallskip
\textbf{2. Medium-impact citations:} These are papers that helped you write your paper, but were not fundamentally irreplaceable. You could have used an alternative prior work or formulation, but you chose this one because it was particularly useful, clear, or canonical.
\begin{itemize}[leftmargin=1.5em, itemsep=0pt, topsep=1pt, parsep=0pt]
    \item \textit{Conceptual or operational contribution (non-unique):} The reference conveys an idea, dataset, or model family that meaningfully helped your setup, but other comparable alternatives exist. Examples: selecting LLaMA-1 vs LLaMA-2; choosing one evaluation protocol among several similar ones; relying on one of several formulations of a known concept.
    \item \textit{Organic helpfulness:} The reference is genuinely helpful for understanding your paper, but not uniquely necessary. It situates your work clearly, but your contribution does not hinge on this specific citation.
    \item \textit{Typical quantity:} roughly 5--15 papers.
\end{itemize}

\smallskip
\textbf{3. Low-impact citations:} These citations provide background, context, or perfunctory acknowledgement, but the core contribution of your paper is not dependent on them in any strong way.
\begin{itemize}[leftmargin=1.5em, itemsep=0pt, topsep=1pt, parsep=0pt]
    \item \textit{Background or definitional citations:} References used to define a task (e.g., Question Answering), introduce a general problem area, or acknowledge standard terminology. The same role could have been fulfilled by many other papers.
    \item \textit{Perfunctory or field-signaling citations:} The reference mainly signals that prior work exists in the broad area. The citing paper does not substantively depend on the specific ideas of the cited work.
    \item \textit{Typical quantity:} the majority of citations.
\end{itemize}

\smallskip
Please output the ranked references as a \textbf{JSON array}, where each entry has:
\begin{itemize}[leftmargin=1.5em, itemsep=0pt, topsep=1pt, parsep=0pt]
    \item \texttt{"rank"}: integer
    \item \texttt{"paperId"}: string
    \item \texttt{"title"}: string
    \item \texttt{"contexts"}: string (all citation contexts)
    \item \texttt{"reason"}: string (why this rank and impactCategory was assigned)
    \item \texttt{"impactCategory"}: string (\texttt{"High"}, \texttt{"Medium"}, \texttt{"Low"})
\end{itemize}

\smallskip
Return \textbf{valid JSON only}, without any extra text. \textbf{DO NOT} wrap the output in \texttt{```} or \texttt{```json}. \textbf{DO NOT} include any explanation, commentary, or text outside the JSON array. \textbf{DO NOT} rank more references than the ones on the list of references given.

\end{tcolorbox}
\caption{Main \methodname{} prompt used in all primary experiments. The model jointly ranks all references $R$ of a given paper $P$ by impact, assigning each to one of three categories (\textit{High}, \textit{Medium}, \textit{Low}). Bracketed fields (\texttt{[main\_paper\_title]}, etc.) are filled at inference time.}
\label{fig:ranking_prompt}
\end{figure*}

\begin{figure*}[t]
\centering
\begin{minipage}{0.95\textwidth} 
\lstset{
    language=json,
    basicstyle=\ttfamily\tiny,   
    numbers=none,                
    breaklines=true,
    frame=single,
    backgroundcolor=\color{gray!5},
    tabsize=2
}
\begin{lstlisting}
[
  {
    "rank": 1,
    "paperId": "5507d267bbf0b4cdb9f893c3c0960a45016f7010",
    "title": "Deep Leakage from Gradients",
    "contexts": "For DLG [1], as described by the authors, we start the procedure with the randomly initialized dummy data and outputs ( x (cid:48), y (cid:48) ), then iteratively update them to minimize the gradient matching objective. | Recent work by Zhu et al. [1] presents an approach (DLG) to steal the proprietary data protected by the participants in distributed learning from the shared gradients. | In this section, we empirically demonstrate the advantages of our (iDLG) method over DLG [1]. | However, recent work by Zhu et al., \"Deep Leakage from Gradient\" (DLG) [1] showed the possibility to steal the private training data from the shared gradients of other participants. | ...for 300 iterations, and evaluate the performance in terms of (i) the accuracy of the extracted labels c (cid:48), and (ii) the fidelity of the extracted Dataset DLG iDLG MNIST 89.9% 100.0% CIFAR-100 83.3% 100.0% LFW 79.1% 100.0% Table 1: Accuracy of the extracted labels for DLG [1] and iDLG. | Following the settings in [1], we use the randomly initialized LeNet for all experiments. | - We empirically demonstrate the advantages of iDLG over DLG [1] via comparing the accuracy of extracted labels and the fidelity of extracted data on three datasets. | This enables us to always extract the ground-truth labels and significantly simplify the objective of DLG [1] in order to extract good-quality data.",
    "reason": "The entire contribution of iDLG is explicitly positioned as an improvement over DLG. The problem setting, core optimization objective (gradient matching), baseline method, experimental protocol (including LeNet and evaluation metrics), and even the paper's name are derived from or defined relative to this work. The logic of iDLG--both conceptually (what it improves) and empirically (what it compares against)--is unintelligible without DLG. Thus it is conceptually and operationally indispensable, making it a high-impact citation.",
    "impactCategory": "High"
  },
  {
    "rank": 2,
    "paperId": "6a6ad9eb495739f4c80e7c09598720c3d5c5dff7",
    "title": "Federated Learning: Collaborative Machine Learning without\nCentralized Training Data",
    "contexts": "In multi-node distributed learning systems such as Collaborative Learning [2, 3, 4] and Federated Learning [5, 6, 7], it is widely believed that sharing gradients between nodes will not leak the private training data.",
    "reason": "This is a canonical paper defining the federated learning paradigm, which provides the main application context where gradient sharing occurs. iDLG's motivation--privacy leakage in distributed / federated setups--relies on this paradigm. However, any of several federated learning introductions could have served a similar role; the iDLG method itself does not technically depend on this specific paper. Hence it is important context but replaceable, so medium impact.",
    "impactCategory": "Medium"
  },
  {
    "rank": 3,
    "paperId": "7fcb90f68529cbfab49f471b54719ded7528d0ef",
    "title": "Federated Learning: Strategies for Improving Communication Efficiency",
    "contexts": "In multi-node distributed learning systems such as Collaborative Learning [2, 3, 4] and Federated Learning [5, 6, 7], it is widely believed that sharing gradients between nodes will not leak the private training data.",
    "reason": "This work further characterizes federated learning systems and communication strategies, reinforcing the setting where gradients are shared. It supports the real-world relevance of the attack scenario but is not directly used in the method or experiments. Other federated learning references could substitute it, so it is context-setting and thus medium rather than high impact.",
    "impactCategory": "Medium"
  },
% omitted entries for brevity
  {
    "rank": 7,
    "paperId": "f2f8f7a2ec1b2ede48cbcd189b376ab9fa0735ef",
    "title": "Privacy-preserving deep learning",
    "contexts": "In multi-node distributed learning systems such as Collaborative Learning [2, 3, 4] and Federated Learning [5, 6, 7], it is widely believed that sharing gradients between nodes will not leak the private training data.",
    "reason": "This paper offers an early privacy-preserving framework for deep learning and is cited to indicate that prior work assumed gradient/parameter sharing can preserve privacy. It supports the contrast between perceived and actual privacy guarantees that iDLG exposes. Nonetheless, the iDLG algorithm does not depend on any of its mechanisms or theory; the citation is mainly motivational and background, so medium impact.",
    "impactCategory": "Medium"
  },
  {
    "rank": 8,
    "paperId": "5d90f06bb70a0a3dced62413346235c02b1aa086",
    "title": "Learning Multiple Layers of Features from Tiny Images",
    "contexts": "We perform experiments on the classification task over three datasets: MNIST [8], CIFAR-100 [9], and LFW [10] with 10, 100, and 5749 categories respectively.",
    "reason": "This technical report defines the CIFAR-100 dataset, which is one of the main benchmarks used to evaluate iDLG's label extraction and data reconstruction performance. The choice of CIFAR-100 helps demonstrate scalability across many classes and more complex images, but in principle another comparable dataset could have been used. It operationally supports experiments but is not uniquely necessary, making it a low-to-medium impact citation; given the categories, it best fits low impact as a standard dataset reference.",
    "impactCategory": "Low"
  },
% omitted entries for brevity
{
    "rank": 11,
    "paperId": "1267fe36b5ece49a9d8f913eb67716a040bbcced",
    "title": "On the limited memory BFGS method for large scale optimization",
    "contexts": "L-BFGS [11] with learning rate 1 is used as the optimizer.",
    "reason": "This optimization paper is referenced to justify the use of L-BFGS for matching gradients when reconstructing data and labels. While L-BFGS may influence convergence behavior in practice, the conceptual contribution of iDLG--the analytic label recovery from gradients--does not depend on this specific optimizer. Many gradient-based optimizers could fill this role with similar effect. Thus it is an implementation-level, replaceable choice and so low impact.",
    "impactCategory": "Low"
  }
]
\end{lstlisting}
\caption{Shows a sample output using Prompt \ref{fig:ranking_prompt} with \textsc{Gpt-5.1}. This is the impact label assignment and ranked references of the citing paper \textit{iDLG: Improved Deep Leakage from Gradients.} 
}
\label{fig:sample_output_prompt}
\end{minipage}
\end{figure*}

\begin{figure*}[t]
  \centering
  \includegraphics[
    width=\textwidth,
    height=\textheight,
    keepaspectratio
  ]{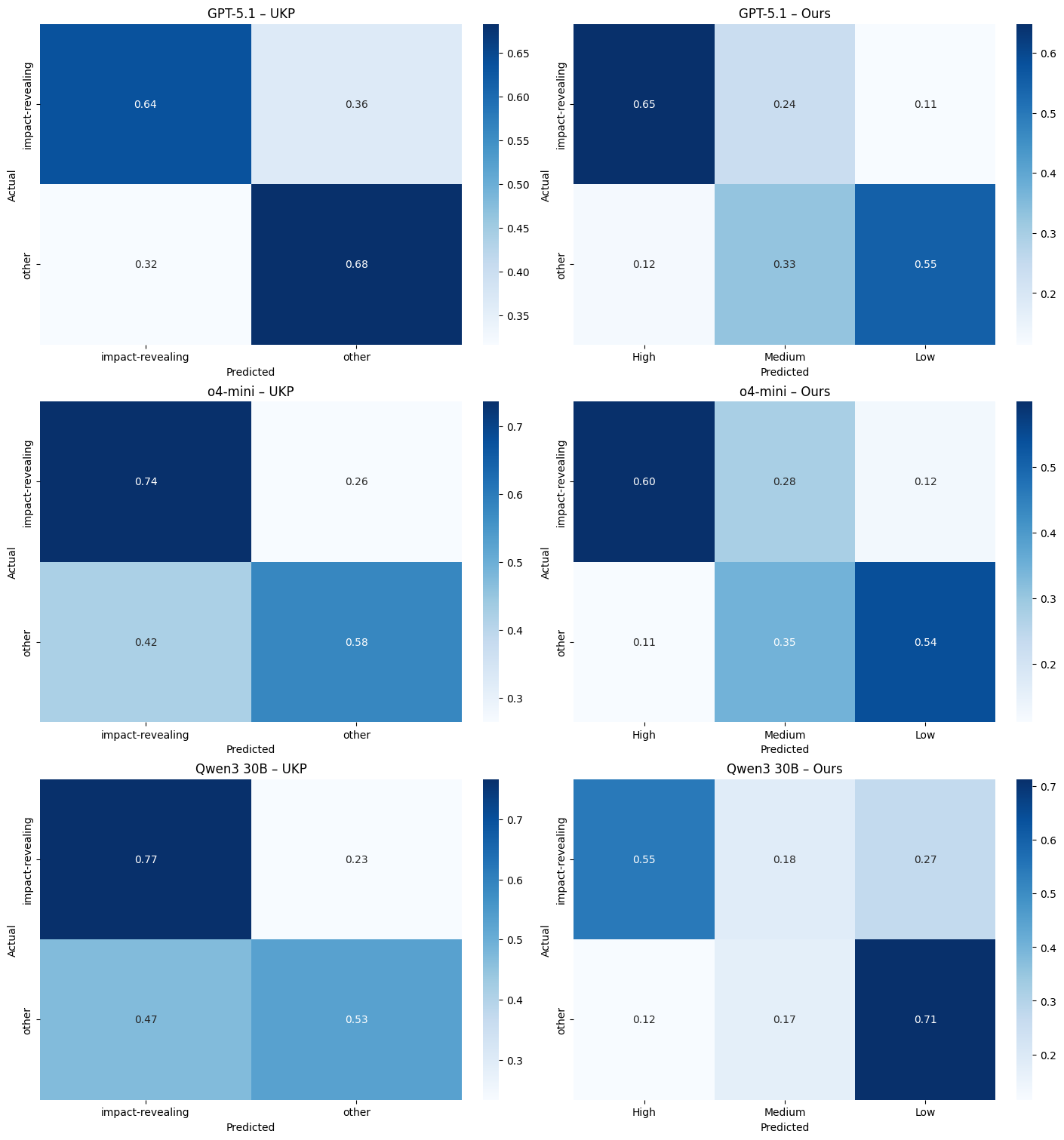}
  \caption{Confusion matrices per model for assigning impact labels through majority vote. Rows represent actual labels; columns represent predicted categories. To compare both approaches (collapsing High to impact-revealing and Medium/Low to other, as described in \ref{EvaluationMeasures}), we find that our method produces more discriminative boundaries, substantially reducing false positives across all three models. Notably, \textsc{Gpt-5.1} achieves this while maintaining comparable recall (0.65 vs. 0.64), demonstrating that our method improves precision without sacrificing sensitivity.}
  \label{fig:confusion_matrices}
\end{figure*}

\section{Results from ablation studies}

\begin{figure*}[t]
  \centering
  \includegraphics[width=0.85\textwidth]{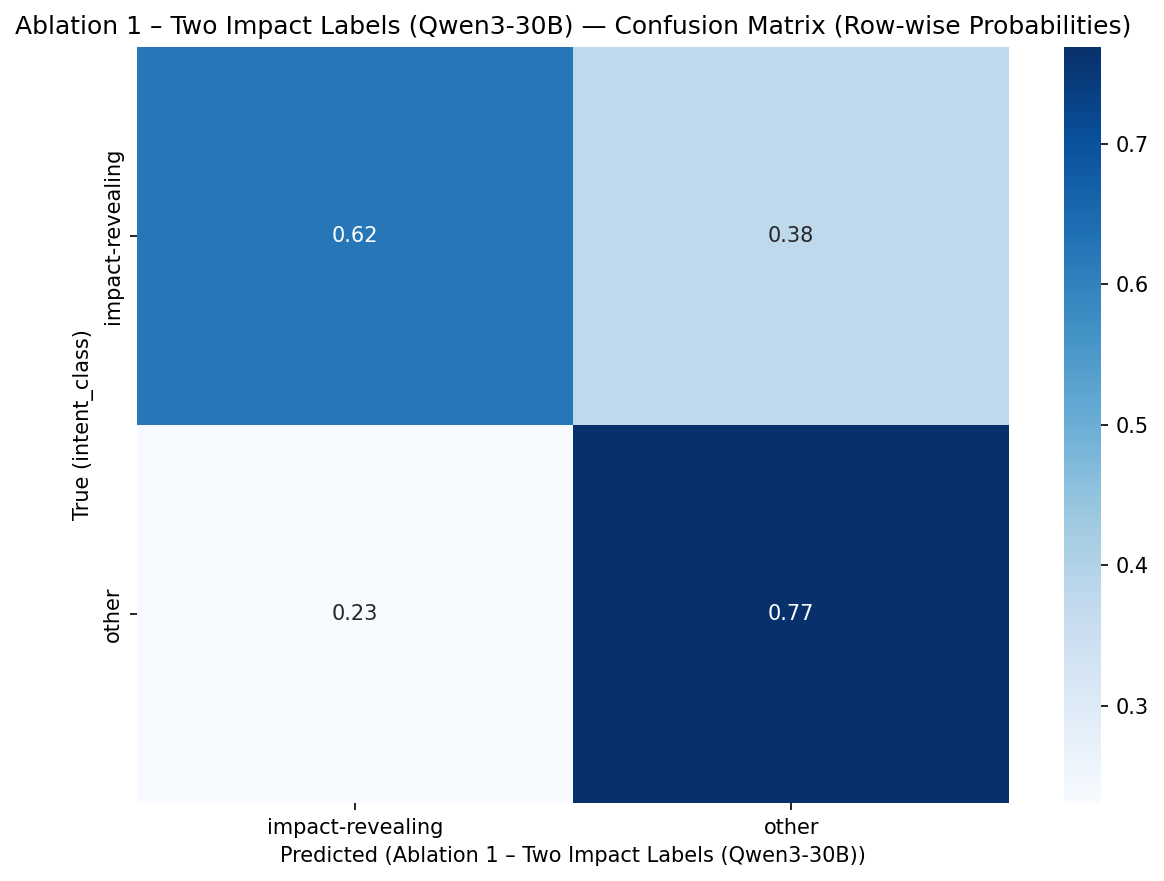}
  \caption{Confusion matrix (row-wise probabilities) for Ablation 1 (two-label variant). Compared to Figure~\ref{fig:confusion_matrices} (Qwen3-30b), the three-label setting introduces a \textit{Medium} bin that captures borderline cases the binary model misclassifies as \textit{impact-revealing}, reducing false positives (precision: $57.1\%\rightarrow70\%$, F1: $+2\%$). Thus, we conclude that three impact labels yield more true impact-revealing classifications and more calibrated judgements.}
  \label{fig:cm_abl1}
\end{figure*}

\begin{figure*}[t]
  \centering
  \includegraphics[width=0.85\textwidth]{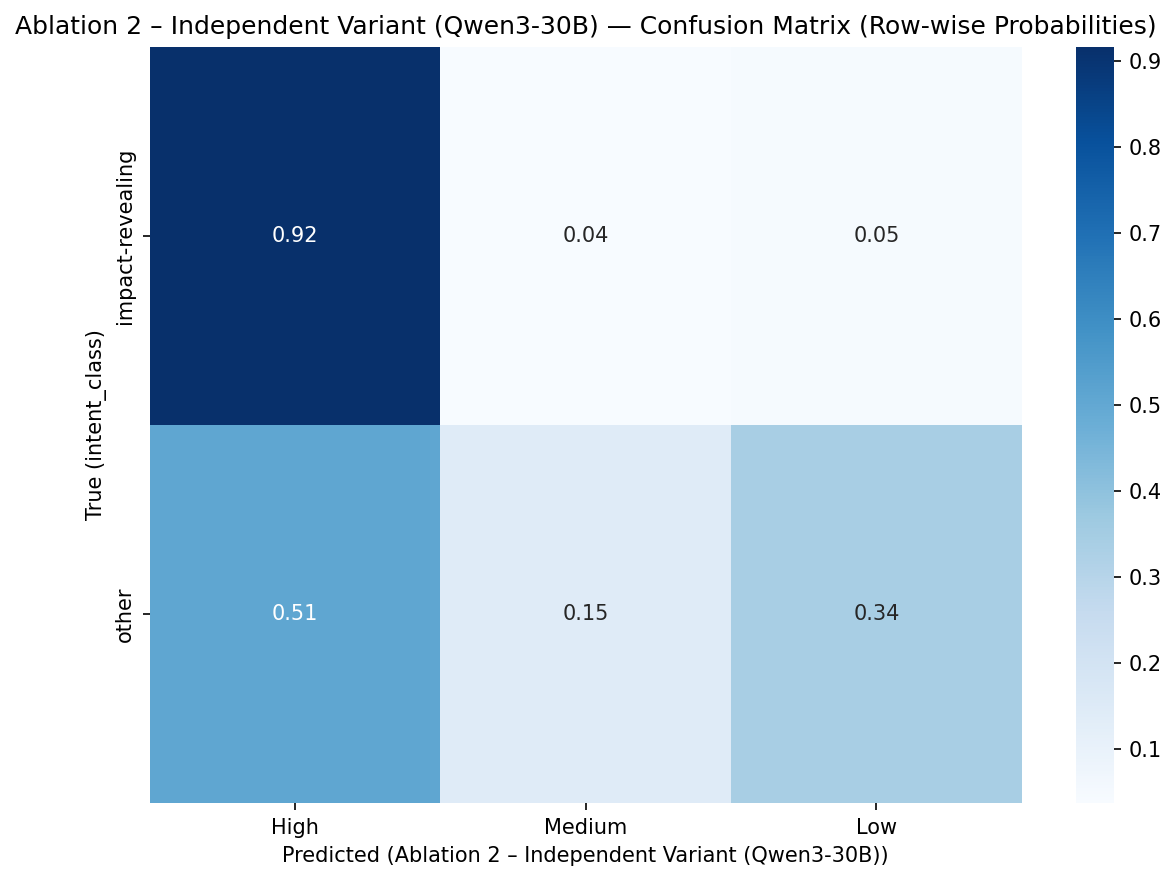}
  \caption{Confusion matrix (row-wise probabilities) for Ablation 2 (independent ranking). The independent variant shows a strong bias toward predicting \textit{High}. In contrast, joint ranking (as shown in Figure~\ref{fig:confusion_matrices}) improves class discrimination and calibration, reducing overprediction and yielding more balanced, reliable predictions.}

  \label{fig:cm_abl2}
\end{figure*}

We perform three ablation studies to isolate the contributions of \methodname{}'s key design choices: label granularity (Table~\ref{table:abl_two_labels}, Figure~\ref{fig:cm_abl1}), joint ranking formulation (Table~\ref{table:abl_independent}, Figure~\ref{fig:cm_abl2}), and permutation-based majority voting (Table~\ref{table:abl3_single_run}).



\renewcommand{\arraystretch}{1.1}
\begin{table*}[t]
\centering
\small
\setlength{\tabcolsep}{8pt}
\definecolor{aclgreen}{RGB}{200,235,200}
\definecolor{aclgray}{RGB}{230,230,230}
\begin{tabular}{l l c c c c c c c}
\toprule
\textbf{Model} & \textbf{Method} & \textbf{Accuracy($\uparrow$)} & \textbf{Bal. Acc.($\uparrow$)} & \textbf{MCC($\uparrow$)} & \textbf{P($\uparrow$)} & \textbf{R($\uparrow$)} & \textbf{F1($\uparrow$)} & \textbf{F1\textsubscript{$-$}($\uparrow$)} \\
\midrule
\multirow{5}{*}{Qwen3}
  & UKP                                          & $60.8{\scriptstyle\pm1.2}$ & 64.9 & 0.28 & 44.5 & 76.7 & 56.3 & 64.5 \\
  & {\methodname} (2 labels)                     & $69.3{\scriptstyle\pm1.1}$ & 68.9 & 0.37 & 57.1 & 60.0 & 58.5 & 78.8 \\
  & \quad $\Delta$\textsubscript{2L$-$UKP}       & \cellcolor{aclgreen}$+8.5$ & \cellcolor{aclgreen}$+4.0$ & \cellcolor{aclgreen}$+0.09$ & \cellcolor{aclgreen}$+12.6$ & $-16.7$ & \cellcolor{aclgreen}$+2.2$ & \cellcolor{aclgreen}$+14.3$ \\
  & \textit{{\methodname} (3 labels)$^\dagger$}  & \textit{$75.1{\scriptstyle\pm1.0}$} & \textit{71.1} & \textit{0.46} & \textit{70.0} & \textit{53.3} & \textit{60.5} & \textit{83.9} \\
  & \quad \textit{$\Delta$\textsubscript{3L$-$2L}$^\dagger$} & \textit{\cellcolor{aclgreen}$+5.8$} & \textit{\cellcolor{aclgreen}$+2.2$} & \textit{\cellcolor{aclgreen}$+0.09$} & \textit{\cellcolor{aclgreen}$+12.9$} & \textit{$-6.7$} & \textit{\cellcolor{aclgreen}$+2.0$} & \textit{\cellcolor{aclgreen}$+5.1$} \\
\bottomrule
\end{tabular}
\caption{Results of Ablation 1, evaluated with Qwen3-30B, ablating {\methodname}'s three-label formulation (high, medium, low) down to two labels (impact-revealing vs.\ other) following \citet{Arnaout2025IndepthRI}. $\Delta_\text{2L$-$UKP}$ shows the gain of {\methodname} (2 labels) over UKP\@. $\Delta_\text{3L$-$2L}$ shows the further gain of three labels over two. \colorbox{aclgreen}{Green} cells indicate positive differences. $^\dagger$Full {\methodname} (3 labels) results shown for reference only. \textbf{The finer granularity of three labels gives the model a more calibrated decision space: the model more reliably identifies truly high-impact citations.}}
\label{table:abl_two_labels}
\end{table*}

\renewcommand{\arraystretch}{1.1}
\begin{table*}[t]
\definecolor{aclgreen}{RGB}{200,235,200}
\centering
\small
\setlength{\tabcolsep}{9pt}
\begin{tabular}{l l c c c c c c c}
\toprule
\textbf{Model} & \textbf{Method} & \textbf{Accuracy($\uparrow$)} & \textbf{Bal. Acc.($\uparrow$)} & \textbf{MCC($\uparrow$)} & \textbf{P($\uparrow$)} & \textbf{R($\uparrow$)} & \textbf{F1($\uparrow$)} & \textbf{F1\textsubscript{$-$}($\uparrow$)} \\
\midrule
\multirow{3}{*}{Qwen3}
 & Independent    & $63.0{\scriptstyle\pm1.1}$ & 70.3 & 0.40 & 46.8 & 91.7  & 61.9 & 63.9 \\
 & \methodname (Joint)   & $75.1{\scriptstyle\pm1.0}$ & 71.1 & 0.46 & 70.0 & 53.3  & 60.5 & 83.9 \\
 & $\Delta$       & \cellcolor{aclgreen}$+12.1$ & \cellcolor{aclgreen}$+0.8$ & \cellcolor{aclgreen}$+0.06$ & \cellcolor{aclgreen}$+23.2$ & $-38.4$ & $-1.4$ & \cellcolor{aclgreen}$+20.0$ \\
\bottomrule
\end{tabular}
\caption{Shows the results of Ablation 2, which compares considering citation contexts \textbf{independently} against \textbf{jointly} ranking of all citations within a paper, evaluated with Qwen3-30B. $\Delta := \text{\methodname} - \text{Independent}$. \colorbox{aclgreen}{Green} cells indicate gains from joint ranking. The precision gains show that considering contexts jointly improves the identification of truly high-impact citations and corrects the overprediction bias observed in the independent setting.}
\label{table:abl_independent}
\end{table*}


\begin{figure*}[p]
\centering
\begin{tcolorbox}[
    colback=gray!5,
    colframe=gray!50,
    boxrule=0.5pt,
    arc=4pt,
    width=\textwidth
]
\small

The task is to rank all of the references $R$ (where $R$ is a list of papers $r_1, r_2, \ldots$) of a given paper $P$ based on how impactful and influential each $r_n$ was on $P$.

\smallskip
\textbf{The paper $P$ you will be analyzing is:}
\begin{itemize}[leftmargin=1.5em, itemsep=0pt, topsep=1pt, parsep=0pt]
    \item Title: \texttt{[main\_paper\_title]}
    \item Abstract: \texttt{[main\_paper\_abstract]}
\end{itemize}

\smallskip
\textbf{The list of references is given below} (each reference includes paper ID, title, abstract, and context of citation): \texttt{[references\_text]}

\smallskip
\textbf{Impact categories:}

\smallskip
\textbf{1. ``impact-revealing'' citations:} These are the papers without which your own work would not have been possible. They supply essential conceptual, methodological, or operational ingredients.
\begin{itemize}[leftmargin=1.5em, itemsep=0pt, topsep=1pt, parsep=0pt]
    \item \textit{Conceptual or operational indispensability:} The reference provides a unique conceptual insight, methodological innovation, dataset, or technique that is directly instrumental to your paper. Examples: a specific algorithm your method extends; a benchmark or dataset your study critically depends on; a theoretical formulation your contribution builds on.
    \item \textit{Organic necessity:} The reference is uniquely and genuinely required for a reader to understand how your paper works or how its core logic unfolds. Without this citation, the intellectual lineage of your method would be opaque or incomplete.
    \item \textit{Typical quantity:} 1--5 papers (or even 1).
\end{itemize}

\smallskip
\textbf{2. ``other'' citations:} These are papers that helped you write your paper, but were not fundamentally irreplaceable. You could have used an alternative prior work or formulation, but you chose this one because it was particularly useful, clear, or canonical. These citations could also provide background, context, or perfunctory acknowledgement, but the core contribution of your paper is not dependent on them in any strong way.
\begin{itemize}[leftmargin=1.5em, itemsep=0pt, topsep=1pt, parsep=0pt]
    \item \textit{Conceptual or operational contribution (non-unique):} The reference conveys an idea, dataset, or model family that meaningfully helped your setup, but other comparable alternatives exist. Examples: selecting LLaMA-1 vs LLaMA-2; choosing one evaluation protocol among several similar ones; relying on one of several formulations of a known concept.
    \item \textit{Organic helpfulness:} The reference is genuinely helpful for understanding your paper, but not uniquely necessary. It situates your work clearly, but your contribution does not hinge on this specific citation.
    \item \textit{Background or definitional citations:} References used to define a task (e.g., Question Answering), introduce a general problem area, or acknowledge standard terminology. The same role could have been fulfilled by many other papers.
    \item \textit{Perfunctory or field-signaling citations:} The reference mainly signals that prior work exists in the broad area. The citing paper does not substantively depend on the specific ideas of the cited work.
    \item \textit{Typical quantity:} the majority of citations.
\end{itemize}

\smallskip
Please output the ranked references as a \textbf{JSON array}, where each entry has:
\begin{itemize}[leftmargin=1.5em, itemsep=0pt, topsep=1pt, parsep=0pt]
    \item \texttt{"rank"}: integer
    \item \texttt{"paperId"}: string
    \item \texttt{"title"}: string
    \item \texttt{"contexts"}: string (all citation contexts)
    \item \texttt{"reason"}: string (why this rank and impactCategory was assigned)
    \item \texttt{"impactCategory"}: string (\texttt{"impact-revealing"}, \texttt{"other"})
\end{itemize}

\smallskip
Return \textbf{valid JSON only}, without any extra text. \textbf{DO NOT} wrap the output in \texttt{```} or \texttt{```json}. \textbf{DO NOT} include any explanation, commentary, or text outside the JSON array. \textbf{DO NOT} rank more references than the ones on the list of references given.

\end{tcolorbox}
\caption{Prompt used in Ablation 1, which collapses \methodname's three-label schema to two labels (impact-revealing vs.\ other) following \citet{Arnaout2025IndepthRI}. Bracketed fields (\texttt{[main\_paper\_title]}, etc.) are filled at inference time.}
\label{fig:prompt_abl_two_labels}
\end{figure*}




\begin{figure*}[p]
\centering
\begin{tcolorbox}[
    colback=gray!5,
    colframe=gray!50,
    boxrule=0.5pt,
    arc=4pt,
    width=\textwidth
]
\small

The task is to assess how impactful and influential a single cited reference was on a given paper $P$.

\smallskip
\textbf{The paper $P$ you will be analyzing is:}
\begin{itemize}[leftmargin=1.5em, itemsep=0pt, topsep=1pt, parsep=0pt]
    \item Title: \texttt{[citing\_title]}
    \item Abstract: \texttt{[citing\_abstract]}
\end{itemize}

\smallskip
\textbf{The cited reference to assess is:}
\begin{itemize}[leftmargin=1.5em, itemsep=0pt, topsep=1pt, parsep=0pt]
    \item Paper ID: \texttt{[cited\_id]}
    \item Title: \texttt{[cited\_title]}
    \item Citation context from $P$: \texttt{[context]}
\end{itemize}

\smallskip
\textbf{Impact categories:}

\smallskip
\textbf{1. High-impact citations:} These are the papers without which your own work would not have been possible. They supply essential conceptual, methodological, or operational ingredients.
\begin{itemize}[leftmargin=1.5em, itemsep=0pt, topsep=1pt, parsep=0pt]
    \item \textit{Conceptual or operational indispensability:} The reference provides a unique conceptual insight, methodological innovation, dataset, or technique that is directly instrumental to your paper. Examples: a specific algorithm your method extends; a benchmark or dataset your study critically depends on; a theoretical formulation your contribution builds on.
    \item \textit{Organic necessity:} The reference is uniquely and genuinely required for a reader to understand how your paper works or how its core logic unfolds. Without this citation, the intellectual lineage of your method would be opaque or incomplete.
    \item \textit{Typical quantity:} 1--5 papers (or even 1).
\end{itemize}

\smallskip
\textbf{2. Medium-impact citations:} These are papers that helped you write your paper, but were not fundamentally irreplaceable. You could have used an alternative prior work or formulation, but you chose this one because it was particularly useful, clear, or canonical.
\begin{itemize}[leftmargin=1.5em, itemsep=0pt, topsep=1pt, parsep=0pt]
    \item \textit{Conceptual or operational contribution (non-unique):} The reference conveys an idea, dataset, or model family that meaningfully helped your setup, but other comparable alternatives exist. Examples: selecting LLaMA-1 vs LLaMA-2; choosing one evaluation protocol among several similar ones; relying on one of several formulations of a known concept.
    \item \textit{Organic helpfulness:} The reference is genuinely helpful for understanding your paper, but not uniquely necessary. It situates your work clearly, but your contribution does not hinge on this specific citation.
    \item \textit{Typical quantity:} roughly 5--15 papers.
\end{itemize}

\smallskip
\textbf{3. Low-impact citations:} These citations provide background, context, or perfunctory acknowledgement, but the core contribution of your paper is not dependent on them in any strong way.
\begin{itemize}[leftmargin=1.5em, itemsep=0pt, topsep=1pt, parsep=0pt]
    \item \textit{Background or definitional citations:} References used to define a task (e.g., Question Answering), introduce a general problem area, or acknowledge standard terminology. The same role could have been fulfilled by many other papers.
    \item \textit{Perfunctory or field-signaling citations:} The reference mainly signals that prior work exists in the broad area. The citing paper does not substantively depend on the specific ideas of the cited work.
    \item \textit{Typical quantity:} the majority of citations.
\end{itemize}

\smallskip
Please output the ranked references as a \textbf{JSON array}, where each entry has:
\begin{itemize}[leftmargin=1.5em, itemsep=0pt, topsep=1pt, parsep=0pt]
    \item \texttt{"paperId"}: string
    \item \texttt{"title"}: string
    \item \texttt{"contexts"}: string (all citation contexts)
    \item \texttt{"reason"}: string (why this rank and impactCategory was assigned)
    \item \texttt{"impactCategory"}: string (\texttt{"High"}, \texttt{"Medium"}, \texttt{"Low"})
\end{itemize}

\smallskip
Return \textbf{valid JSON only}, without any extra text. \textbf{DO NOT} wrap the output in \texttt{```} or \texttt{```json}. \textbf{DO NOT} include any explanation, commentary, or text outside the JSON array. \textbf{DO NOT} rank more references than the ones on the list of references given.

\end{tcolorbox}
\caption{Prompt used in Ablation 2, which scores each citation context \textbf{independently} (i.e., in isolation, without access to the full reference list). The model is queried three times per citation and the majority label is taken as the final prediction. Bracketed fields (\texttt{[citing\_title]}, etc.) are filled at inference time.}
\label{fig:prompt_abl_independent}
\end{figure*}

\renewcommand{\arraystretch}{1.1}
\begin{table*}[t]
\definecolor{aclgreen}{RGB}{200,235,200}
\centering
\small
\setlength{\tabcolsep}{10pt}
\begin{tabular}{l l c c c c c c c}
\toprule
\textbf{Model} & \textbf{Method} & \textbf{Accuracy($\uparrow$)} & \textbf{Bal. Acc.($\uparrow$)} & \textbf{MCC($\uparrow$)} & \textbf{P($\uparrow$)} & \textbf{R($\uparrow$)} & \textbf{F1($\uparrow$)} & \textbf{F1\textsubscript{$-$}($\uparrow$)} \\
\midrule
\multirow{5}{*}{\textsc{Gpt-5.1}}
 & Run 1          & $76.8{\scriptstyle\pm1.0}$ & 75.3 & 0.53 & 71.7 & 62.7  & 66.9 & 85.2 \\
 & Run 2          & $77.1{\scriptstyle\pm1.0}$ & 75.3 & 0.52 & 71.1 & 63.2  & 66.9 & 85.1 \\
 & Run 3          & $76.9{\scriptstyle\pm1.0}$ & 75.7 & 0.53 & 71.1 & 64.2  & 67.5 & 85.2 \\
 & \methodname    & $78.6{\scriptstyle\pm1.0}$ & 75.8 & 0.54 & 72.2 & 63.7  & \textbf{67.7} & 85.5 \\
 & $\Delta$       & \cellcolor{aclgreen}$+1.5$ & \cellcolor{aclgreen}$+0.1$ & \cellcolor{aclgreen}$+0.01$ & \cellcolor{aclgreen}$+0.5$ & $-0.5$ & \cellcolor{aclgreen}$+0.2$ & \cellcolor{aclgreen}$+0.3$ \\
\midrule
\multirow{5}{*}{o4-mini}
 & Run 1          & $53.5{\scriptstyle\pm1.2}$ & 72.4 & 0.50 & 76.0 & 53.0  & 62.5 & 85.5 \\
 & Run 2          & $52.7{\scriptstyle\pm1.2}$ & 72.3 & 0.50 & 77.3 & 52.1  & 62.3 & 85.6 \\
 & Run 3          & $52.6{\scriptstyle\pm1.2}$ & 71.8 & 0.48 & 74.5 & 52.3  & 61.5 & 85.0 \\
 & \methodname    & $65.4{\scriptstyle\pm1.1}$ & 74.2 & 0.53 & 76.3 & 57.1  & 65.3 & 86.0 \\
 & $\Delta$       & \cellcolor{aclgreen}$+11.9$ & \cellcolor{aclgreen}$+1.8$ & \cellcolor{aclgreen}$+0.03$ & $-1.0$ & \cellcolor{aclgreen}$+4.1$ & \cellcolor{aclgreen}$+2.8$ & \cellcolor{aclgreen}$+0.4$ \\
\midrule
\multirow{5}{*}{Qwen3}
 & Run 1          & $65.3{\scriptstyle\pm1.1}$ & 68.3 & 0.40 & 66.9 & 48.2  & 56.0 & 82.6 \\
 & Run 2          & $66.0{\scriptstyle\pm1.1}$ & 70.1 & 0.46 & 73.1 & 49.1  & 58.7 & 84.3 \\
 & Run 3          & $66.6{\scriptstyle\pm1.1}$ & 69.1 & 0.43 & 69.7 & 48.6  & 57.2 & 83.4 \\
 & \methodname    & $75.1{\scriptstyle\pm1.0}$ & 71.1 & 0.46 & 70.0 & 53.3  & 60.5 & 83.9 \\
 & $\Delta$       & \cellcolor{aclgreen}$+8.5$ & \cellcolor{aclgreen}$+1.0$ & $0.00$ & $-3.1$ & \cellcolor{aclgreen}$+4.2$ & \cellcolor{aclgreen}$+1.8$ & $-0.4$ \\
\bottomrule
\end{tabular}
\caption{Shows the results of Ablation 3, which studies the effect of \textbf{majority voting} across three randomized orderings vs.\ any single run. $\Delta := \text{\methodname} - \text{best single run}$, where the best is taken column-wise across the three runs. \colorbox{aclgreen}{Green} cells show gains from majority voting. \textbf{Bold} indicates the highest F1 overall. \textbf{Majority voting consistently matches or exceeds even the strongest individual run}, demonstrating robustness to positional bias.}
\label{table:abl3_single_run}
\end{table*}

\section{Analysis of Missing References}
\label{Sec:Missing References}
\begin{figure*}[t]
  \centering
  \includegraphics[
    width=\textwidth,
    height=\textheight,
    keepaspectratio
  ]{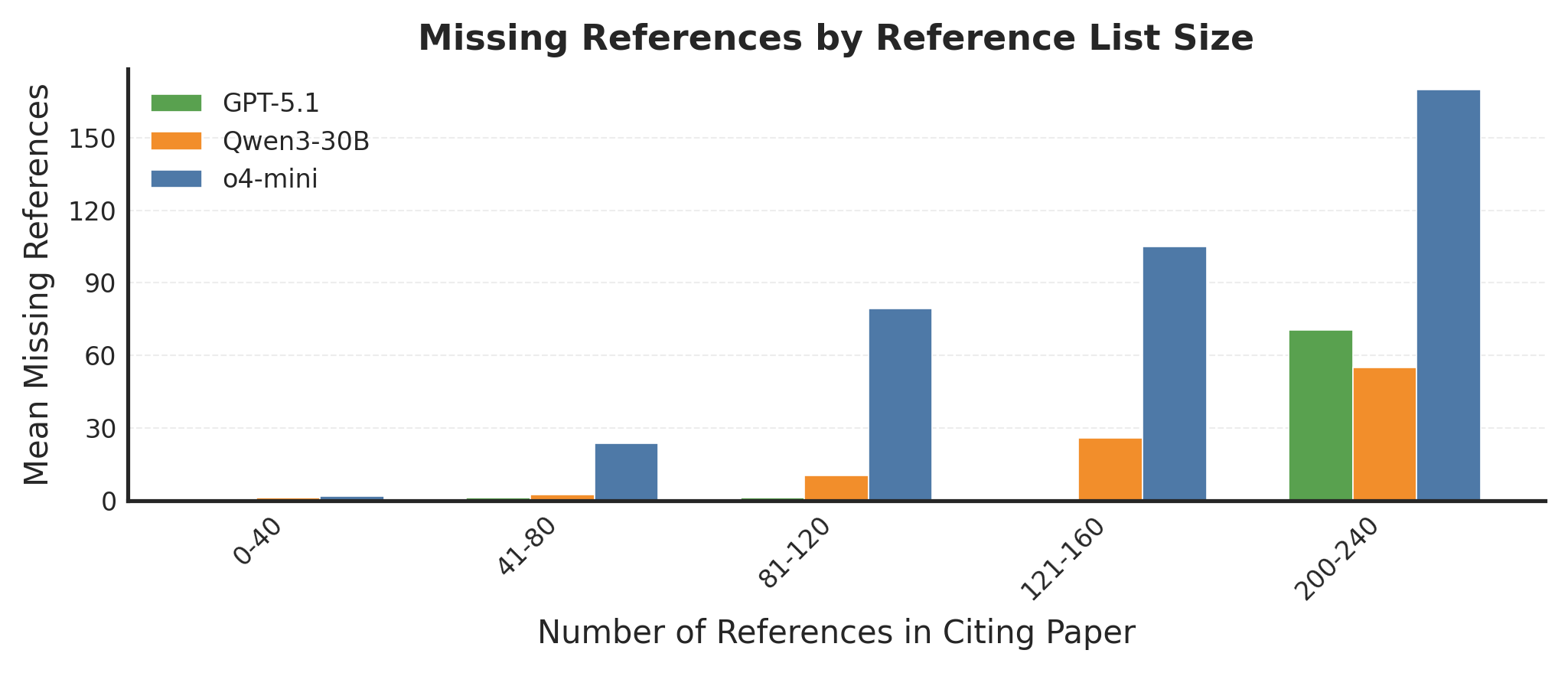}
  \caption{Mean number of missing references in the aggregated ranks per model as a function of the number of references in the citing paper. All three models rank nearly all references when the reference list is less than 40. However, they omit more references as the list grows longer. o4-mini exhibits the steepest degradation, while \textsc{Gpt-5.1} and Qwen3-30B are more robust.}
  \label{fig:missing_refs}
\end{figure*}
As noted in \S\ref{Sec:Experiments}, we rank references by their impact on the citing paper. We find that as the number of references grows, models struggle to rank the complete list despite having large context windows of at least \(200\text{k}\) tokens. Figure~\ref{fig:missing_refs} shows this effect. 

When the citing paper contains fewer than 40 references, all three models successfully rank nearly every reference. However, as the reference list grows, the number of missing references increases sharply. \textsc{Gpt-5.1} and Qwen3-30B degrade more gracefully, with \textsc{Gpt-5.1} omitting roughly 70 references on average for papers with 200--240 references. In contrast, o4-mini exhibits the steepest decline, failing to rank over 170 references on average in the same range. 

These results suggest that, although the models can technically ingest long contexts, they struggle to rank all items as the reference list grows.

\section{Qualitative Error Analysis Details}
\label{app:error-analysis}

We provide additional material supporting the qualitative error analysis in \S\ref{sec:error-analysis}. We include examples of both false positives (FP) and false negatives (FN), drawn from cases that were unanimously misclassified by all three models (Qwen3-30B, \textsc{Gpt-5.1}, and o4-mini) under \methodname{}, identifying commonalities in these errors.

\begin{figure*}[t]
\centering
\includegraphics[width=\linewidth]{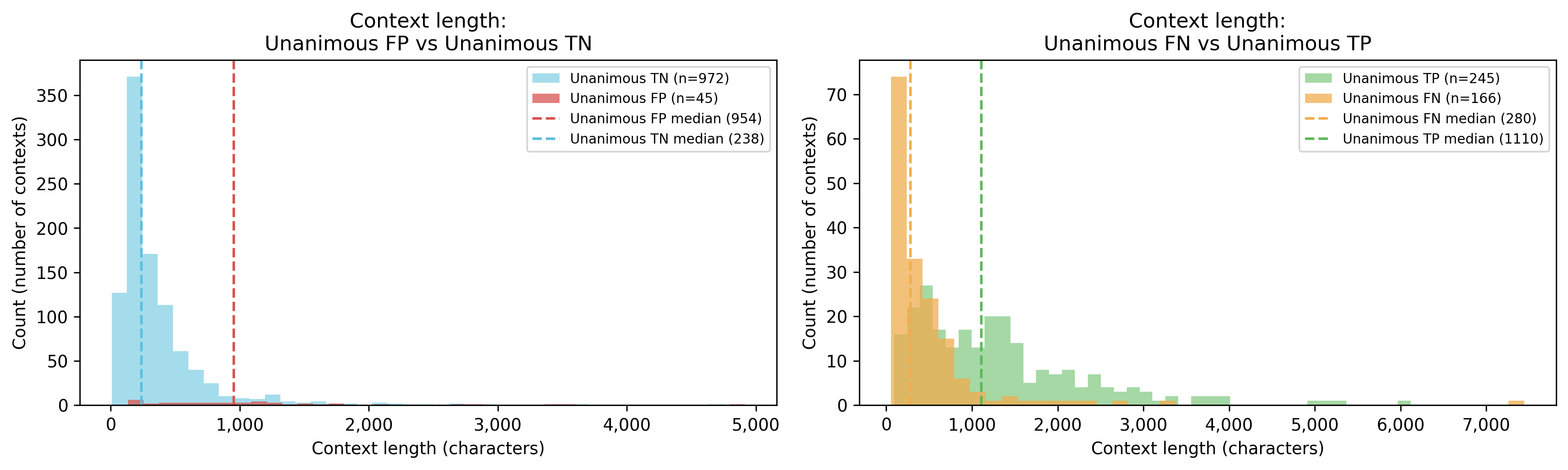}
\caption{Distribution of citation context length (in characters) for cases unanimously classified by \textsc{Qwen3-30B}, \textsc{Gpt-5.1}, and \textsc{o4-mini} under \methodname{}. \textbf{Left:} unanimous false positives (FP, n=45) versus unanimous true negatives (TN, n=972), with medians of 954 and 238 characters respectively. \textbf{Right:} unanimous false negatives (FN, n=166) versus unanimous true positives (TP, n=245), with medians of 280 and 1{,}110 characters respectively. Dashed lines mark the medians for each group.}
\label{fig:length-hist}
\end{figure*}

Figure~\ref{fig:length-hist} compares context lengths between unanimously misclassified cases and correctly classified instances. Figures~\ref{fig:fp-examples} and~\ref{fig:fn-examples} present representative examples for each error type.

\begin{figure*}[p]
\centering
\begin{tcolorbox}[
colback=gray!5,
colframe=gray!50,
boxrule=0.5pt,
arc=4pt,
width=\textwidth
]
\small
To automatically obtain a large training dataset, DS has been proposed (Mintz et al., 2009). Mintz et al.\ (2009) reports that distant supervision may lead to more than 30
\smallskip
\hrule
\smallskip
(Ren et al.\ 2016) proposed a multimodal Long Short-Term Memory (LSTM) for speaker identification, which referred to locating a person who has the same identity with the ongoing sound in a certain video. Inspired by (Ren et al.\ 2016), we propose to build temporal dependency across visual and audio modalities through sharing weights for video caption, aiming at exploring whether temporal dependency across visual and audio modalities can capture the resonance information among them or not. Ren et al.\ (Ren et al.\ 2016) proposed a multimodal Long Short-Term Memory (LSTM) for speaker identification, which referred to locating a person who has the same identity with the ongoing sound in a certain video.
\smallskip
\hrule
\smallskip
More recent studies (Brock et al., 2018; Shirakawa et al., 2018; Pham et al., 2018; Liu et al., 2019; Xie et al., 2019; Cai et al., 2019), on the other hand, optimize the weights and the architecture simultaneously within a single training by treating all possible architectures as subgraphs of a supergraph. We generalize the work by Shirakawa et al.\ (2018) to enable arbitrary types of architecture variables including categorical variables, ordinal (such as real or integer) variables, and their mixture. In the last direction, promising approaches transform a coupled optimization of weights and architectures into optimization of a differentiable objective by means of continuous relaxation (Liu et al., 2019; Xie et al., 2019) or stochastic relaxation (Shirakawa et al., 2018; Pham et al., 2018). \dots Shirakawa et al.\ (2018) has introduced it to model connections and types of activation functions in multi-layer perceptrons.
\end{tcolorbox}
\caption{Examples of contexts that were misclassified as impact-revealing by \textsc{Qwen3-30B}, \textsc{Gpt-5.1}, and \textsc{o4-mini} under \methodname{}. In each case the cited work is described in detail and attributed to its authors, although the citing paper does not build on it.}
\label{fig:fp-examples}
\end{figure*}
\begin{figure*}[p]
\centering
\begin{tcolorbox}[
colback=gray!5,
colframe=gray!50,
boxrule=0.5pt,
arc=4pt,
width=\textwidth
]
\small
[11] leveraged the reinforcement learning to automatically prune the convolution channels.
\smallskip
\hrule
\smallskip
Our method also connects to PinSage [21] and GAT [15].
\smallskip
\hrule
\smallskip
We run the 4 models on the datasets and respective splits from [Yang et al., 2016].
\smallskip
\hrule
\smallskip
Discriminative models are known to suffer from catastrophic forgetting when learning sequentially from examples from a single class at a time, and specialized techniques are actively being developed to minimize this problem (Rusu et al., 2016; Kirkpatrick et al., 2017; Fernando et al., 2017).
\smallskip
\hrule
\smallskip
We use a baseline Tacotron architecture specified in [8], where we use a GMM attention [9], LSTM-based decoder with zoneout regularization [10] and phoneme inputs derived from normalized text.
\end{tcolorbox}
\caption{Examples of contexts that were misclassified as not impactful by \textsc{Qwen3-30B}, \textsc{Gpt-5.1}, and \textsc{o4-mini} under \methodname{}. The citing paper builds on or is inspired by the cited work, but the mention is brief or the cited paper is grouped with other references in the same citation.}
\label{fig:fn-examples}
\end{figure*}

\section{Alternative Impact Label Assignment Approach}
\label{Sec:AlternativeApproach}
While majority voting performs strongly as an aggregation technique for impact classification (\S\ref{section:method}), a limitation of this approach is that the resulting labels are not guaranteed to decrease monotonically with rank. 

For example, the paper ranked fifth could receive a \textit{Low} label, while the paper ranked sixth could receive a \textit{Medium} label. This violates the assumption that impact categories follow the order \textit{High} $>$ \textit{Medium} $>$ \textit{Low}.

To address cases where an aggregated ranking with references sorted by impact is needed, we propose an alternative method based on ordinal regression that ensures labels respect this ordering.

\noindent\textbf{Aggregating Ranks:} 
\label{aggregating-ranks}

We aggregate the three rankings generated by the LLM judge for each citing paper using Reciprocal Rank Fusion (RRF) \cite{cormack2009reciprocal}. Following \citet{cormack2009reciprocal}, we compute the RRF score as
\[
\text{RRF-score}(p) = \sum_{i=1}^{N} \frac{1}{k + \text{rank}_i(p)},
\]
where \( p \) denotes a cited paper, \( N = 3 \) since each citing paper produces three rankings, \( k \) is a constant (we set \( k = 60 \)), and \( \text{rank}_i(p) \) is the rank of paper \( p \) in the \( i \)-th ranking.

As noted in \S\ref{Sec:Missing References}, some references do not appear in all ranked lists. For such cases, we compute the mean of the non-empty values among \( \text{rank}_1(p) \), \( \text{rank}_2(p) \), and \( \text{rank}_3(p) \), and use this value to impute the missing ranks. If a paper does not appear in any of the three rankings, it is excluded from the final aggregated list. In addition, hallucinated references that are not present in the original reference list are discarded during aggregation. Furthermore, if a cited paper appears multiple times within the same ranking, we retain only its lowest (best) rank. For example, if a paper \( p \) is ranked both 1 and 3 in the same ranking \( r_i \), we use 1 as \( \text{rank}_i(p) \).

We release the dataset of aggregated rankings produced by the LLM judges using the models \textsc{Gpt-5.1}, o4-mini, and Qwen3-30B-A3B-Instruct-2507-FP8 \cite{qwen3technicalreport}. Specifically, the dataset contains the aggregated rankings for each of the 442 citing papers per model we generated in our experiments (\S\ref{Sec:Experiments}), for a total of 1,326 files. A sample aggregated ranking is shown in Figure~\ref{fig:sample_output_ord_reg}. 

\noindent\textbf{Predicting Impact Labels with an Ordinal Regression Model:}

\noindent\textbf{Training Data Construction.}
For each of the 442 citing papers, we collect the ranked reference lists produced by the three independent LLM runs described in \S\ref{section:method}. For each cited reference, we construct a feature vector consisting of the raw ranks from each run ($r_1, r_2, r_3$), the normalized ranks ($r_i / N$, where $N$ is the total number of references in the citing paper), the standard deviation across the three runs, and the mean rank. If a reference is missing a rank from one or two runs, we impute the missing value(s) with the median of the available ranks. Normalizing by the reference list length ensures comparability across citing papers with different numbers of references. The impact label (\textit{Low}, \textit{Medium}, or \textit{High}) for the training data is determined by majority vote over the three runs' impact category assignments.

To prevent data leakage, we exclude all (citing paper, cited paper) pairs that appear in our held-out test set prior to model fitting. The held-out test set consists of the 1,338 cited papers for which we have ground-truth impact annotations from the dataset released by \citet{Arnaout2025IndepthRI}. Therefore, the training data comprises only the remaining references that were ranked alongside the test pairs during the per-paper ranking step but do not themselves appear in the evaluation set. Importantly, the normalization factor $N$ is computed from the full reference list before exclusion, preserving the original scale of the citing paper.

\noindent\textbf{Model Training.} We train a single global ordinal regression model pooled across all citing papers, rather than fitting per-paper models, to maximize training signal. We use the Immediate-Threshold variant of ordinal logistic regression \citep{pedregosa2015feature} from the \texttt{mord} library, which respects the natural ordering of the three impact categories through shared threshold parameters. The model is regularized with an L2 penalty ($\alpha = 1.0$).

\noindent\textbf{Prediction.}
\renewcommand{\arraystretch}{1.2}
\begin{table}[t]
\centering
\small
\setlength{\tabcolsep}{6pt}

\definecolor{aclgray}{gray}{0.92}

\resizebox{\columnwidth}{!}{%
\begin{tabular}{>{\centering\arraybackslash}p{2.3cm}
                >{\raggedright\arraybackslash}p{2.0cm}
                c c c c}
\toprule

\textbf{Citation Impact Classifier} &
\textbf{Model} &
\textbf{Acc.($\uparrow$)} & \textbf{P($\uparrow$)} & \textbf{R($\uparrow$)} & \textbf{F1($\uparrow$)} \\

\midrule

\multirow{3}{*}{\textbf{Ours (Ord. Reg.)}}
& \multicolumn{1}{>{\columncolor{aclgray}\raggedright\arraybackslash}p{2.0cm}}{\textsc{Gpt-5.1}}
& \multicolumn{1}{>{\columncolor{aclgray}\centering\arraybackslash}c}{\boldmath{$78.9 {\scriptstyle \pm 1}$}}
& \multicolumn{1}{>{\columncolor{aclgray}\centering\arraybackslash}c}{73.6}
& \multicolumn{1}{>{\columncolor{aclgray}\centering\arraybackslash}c}{62.7}
& \multicolumn{1}{>{\columncolor{aclgray}\centering\arraybackslash}c}{\textbf{67.7}} \\

& o4-mini   & {$64.3 {\scriptstyle \pm 1.1}$} & 76.4 & 54.5 & 62.2 \\
& Qwen3 30B & {$75.9 {\scriptstyle \pm 1}$} & 75.6 & 48.6 & 59.1 \\

\bottomrule
\end{tabular}%
}

\caption{Results using an ordinal regression model to predict impact labels. Reported accuracies include the standard error ($\pm$ SE).}
\label{table:ordinal_regression}
\end{table}
At inference time, we apply the trained global model to the aggregated ranking produced by RRF (\S\ref{aggregating-ranks}) for each citing paper. For every reference in the aggregated list, we construct the same feature vector used during training from its per-run ranks and predict an impact category (\textit{Low}, \textit{Medium}, or \textit{High}). The predicted label is then assigned to the corresponding entry in the RRF-aggregated ranking, yielding a final output in which each cited reference has both an aggregated rank and a predicted impact category. A sample output is shown in Figure~\ref{fig:sample_output_ord_reg}.

\begin{figure*}[t]
  \centering
  \includegraphics[
    width=\textwidth,
    height=\textheight,
    keepaspectratio
  ]{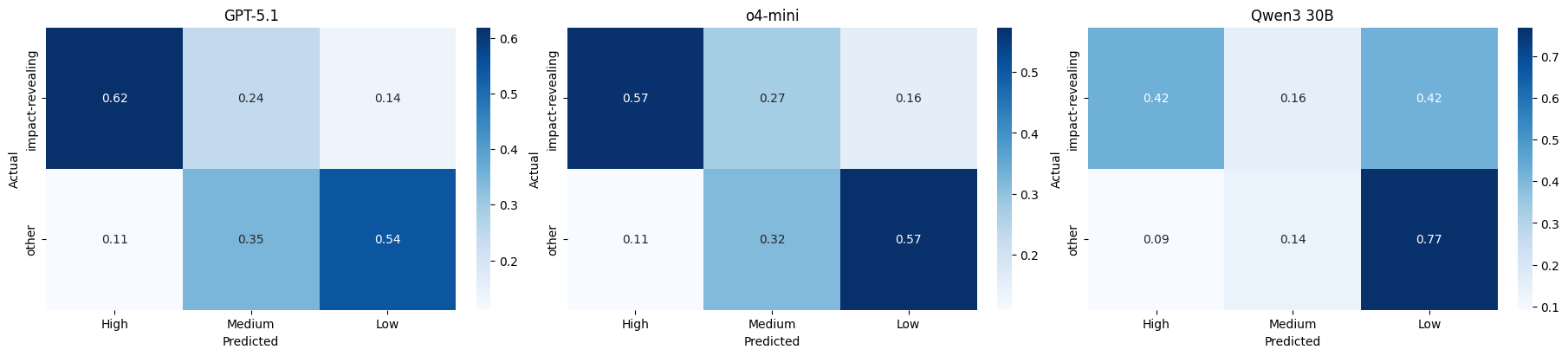}
  \caption{Confusion matrices per model for assigning impact labels with an ordinal regression model. Rows represent actual labels; columns represent predicted categories. \textsc{Gpt-5.1} achieves the best performance in identifying impactful citations and reducing false positive. However, we see a slight decrease in performance with Qwen3-30b. }
  \label{fig:confusion_matrices_ordinal_regression}
\end{figure*}

\noindent\textbf{Results.} Table~\ref{table:ordinal_regression} reports the performance of the ordinal regression model. These results are evaluating the same 1,338 cited papers described in \S\ref{Sec:Experiments}. Our method outperforms the approach of \citet{Arnaout2025IndepthRI} across all models. Compared to our majority-vote approach (Table~\ref{table:full_results}), ordinal regression yields similar overall performance: \textsc{Gpt-5.1} improves slightly in accuracy (78.9 vs.\ 78.6) while maintaining the same F1 of 67.7. For o4-mini and Qwen3-30B, the model trades recall for higher precision, resulting in slightly lower F1 scores.  These results suggest that ordinal regression provides a lightweight post-processing refinement over majority vote without requiring additional LLM calls, with the largest benefit when the underlying LLM judge is already strong. The qualitative results are shown in Figure~\ref{fig:confusion_matrices_ordinal_regression}.

\begin{figure*}[t]
\centering
\begin{minipage}{0.95\textwidth} 
\lstset{
    language=json,
    basicstyle=\ttfamily\tiny,   
    numbers=none,                
    breaklines=true,
    frame=single,
    backgroundcolor=\color{gray!5},
    tabsize=2
}
\begin{lstlisting}
[
  {
    "rank": 1,
    "paperId": "5507d267bbf0b4cdb9f893c3c0960a45016f7010",
    "title": "Deep Leakage from Gradients",
    "rrf_score": 0.04918032786885246,
    "num_rankings_found": 3,
    "predicted_impact": "High"
  },
  {
    "rank": 2,
    "paperId": "6a6ad9eb495739f4c80e7c09598720c3d5c5dff7",
    "title": "Federated Learning: Collaborative Machine Learning without\nCentralized Training Data",
    "rrf_score": 0.04788306451612903,
    "num_rankings_found": 3,
    "predicted_impact": "Medium"
  },
  {
    "rank": 3,
    "paperId": "7fcb90f68529cbfab49f471b54719ded7528d0ef",
    "title": "Federated Learning: Strategies for Improving Communication Efficiency",
    "rrf_score": 0.047619047619047616,
    "num_rankings_found": 3,
    "predicted_impact": "Medium"
  },
  {
    "rank": 4,
    "paperId": "8a564ee07fa930ebc1176019deacdc9951063a99",
    "title": "Collaborative Learning for Deep Neural Networks",
    "rrf_score": 0.046153846153846156,
    "num_rankings_found": 3,
    "predicted_impact": "Medium"
  },
  {
    "rank": 5,
    "paperId": "49bdeb07b045dd77f0bfe2b44436608770235a23",
    "title": "Federated Learning: Challenges, Methods, and Future Directions",
    "rrf_score": 0.04595588235294118,
    "num_rankings_found": 3,
    "predicted_impact": "Medium"
  },
  {
    "rank": 6,
    "paperId": "8bdf6f03bde08c424c214188b35be8b2dec7cdea",
    "title": "Inference Attacks Against Collaborative Learning",
    "rrf_score": 0.045228403437358664,
    "num_rankings_found": 3,
    "predicted_impact": "Medium"
  },
  {
    "rank": 7,
    "paperId": "f2f8f7a2ec1b2ede48cbcd189b376ab9fa0735ef",
    "title": "Privacy-preserving deep learning",
    "rrf_score": 0.04500226142017187,
    "num_rankings_found": 3,
    "predicted_impact": "Medium"
  },
  {
    "rank": 8,
    "paperId": "1267fe36b5ece49a9d8f913eb67716a040bbcced",
    "title": "On the limited memory BFGS method for large scale optimization",
    "rrf_score": 0.04429804634257156,
    "num_rankings_found": 3,
    "predicted_impact": "Low"
  },
  {
    "rank": 9,
    "paperId": "5d90f06bb70a0a3dced62413346235c02b1aa086",
    "title": "Learning Multiple Layers of Features from Tiny Images",
    "rrf_score": 0.04390451832907076,
    "num_rankings_found": 3,
    "predicted_impact": "Low"
  },
  {
    "rank": 10,
    "paperId": "c6b3ca4f939e36a9679a70e14ce8b1bbbc5618f3",
    "title": "Labeled Faces in the Wild: A Database forStudying Face Recognition in Unconstrained Environments",
    "rrf_score": 0.0432712215320911,
    "num_rankings_found": 3,
    "predicted_impact": "Low"
  },
  {
    "rank": 11,
    "paperId": "162d958ff885f1462aeda91cd72582323fd6a1f4",
    "title": "Gradient-based learning applied to document recognition",
    "rrf_score": 0.04265593561368209,
    "num_rankings_found": 3,
    "predicted_impact": "Low"
  }
]

\end{lstlisting}
\caption{Shows a sample output of the ordinal regression model predicting impact labels for the ranked references of the citing paper \textit{iDLG: Improved Deep Leakage from Gradients}. 
}
\label{fig:sample_output_ord_reg}
\end{minipage}
\end{figure*}

\section{Pilot Study}
\label{sec:pilot_study}

Prior to running the experiments described in \S\ref{Sec:Experiments}, we conducted a pilot study to evaluate the effectiveness of the prompt shown in Figure~\ref{fig:ranking_prompt} for ranking references by their impact on their citing papers. 
Using a custom annotation interface we designed, six annotators ranked references from a paper they co-authored according to our predefined impact criteria. The annotators were PhD students from our lab who volunteered to participate; no monetary compensation was offered and all names are anonymized.

\begin{figure*}[t]
  \centering
  \includegraphics[width=\textwidth]{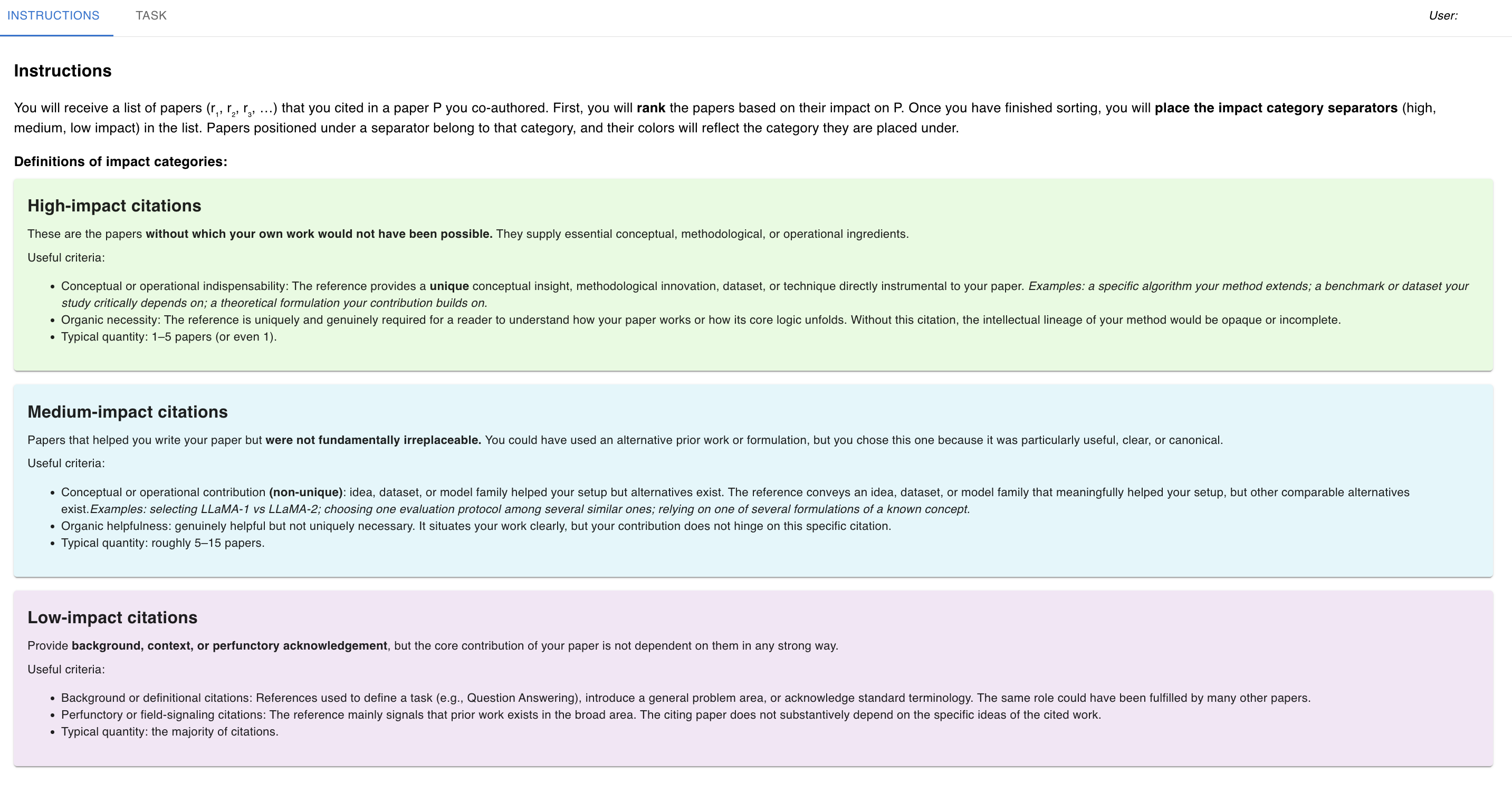}
  \caption{The main layout of the custom annotation interface used in the pilot study, showing the overall task setup and instructions for ranking references by impact.}
  \label{fig:Pilot_1}
\end{figure*}

\begin{figure*}[t]
  \centering
  \includegraphics[width=\textwidth]{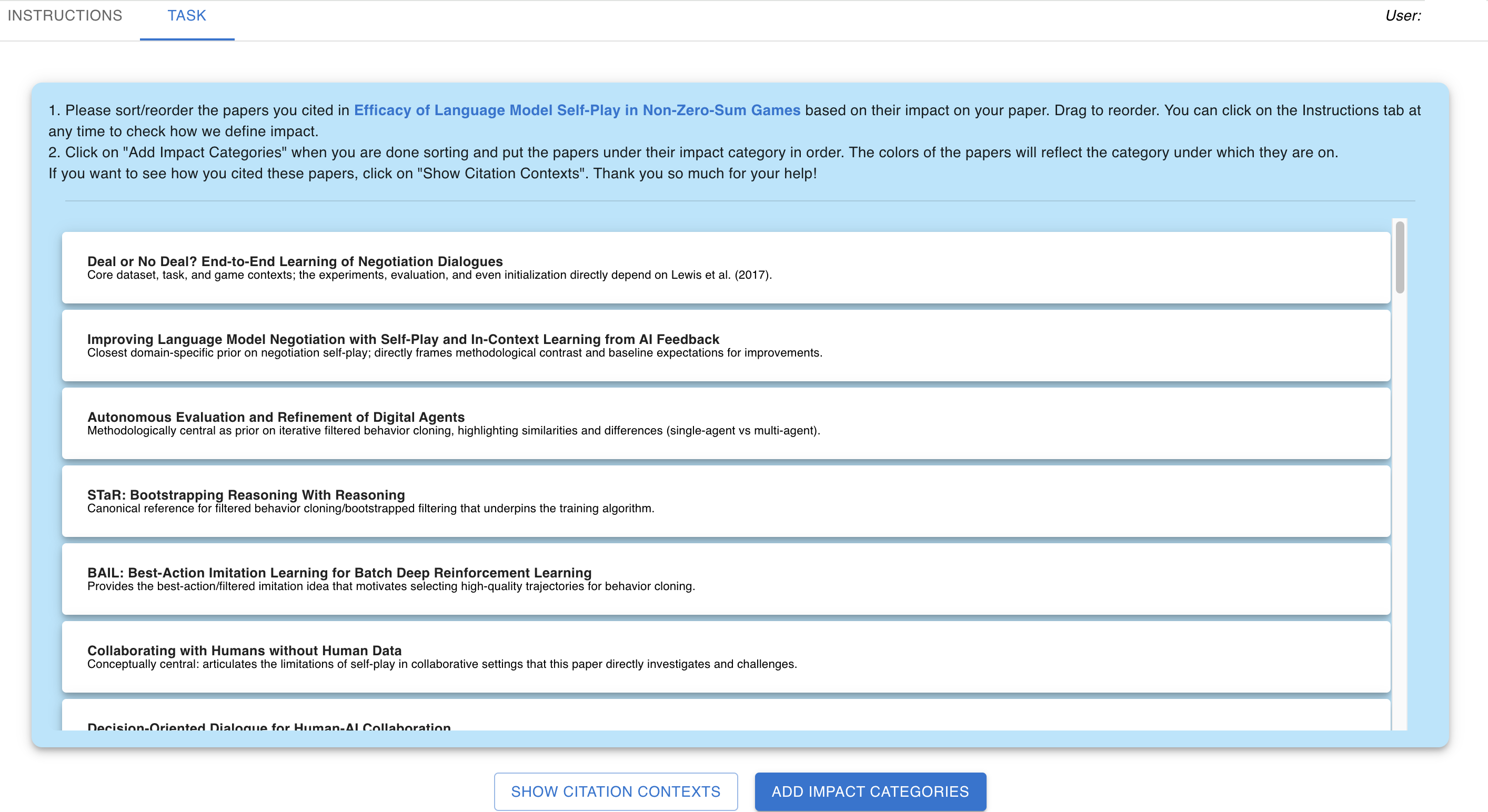}
  \caption{The full list of references from an annotator's paper displayed within the interface, allowing users to rank their references by dragging and reordering them according to the impact criteria.}
  \label{fig:Pilot_2}
\end{figure*}

\begin{figure*}[t]
  \centering
  \includegraphics[width=\textwidth]{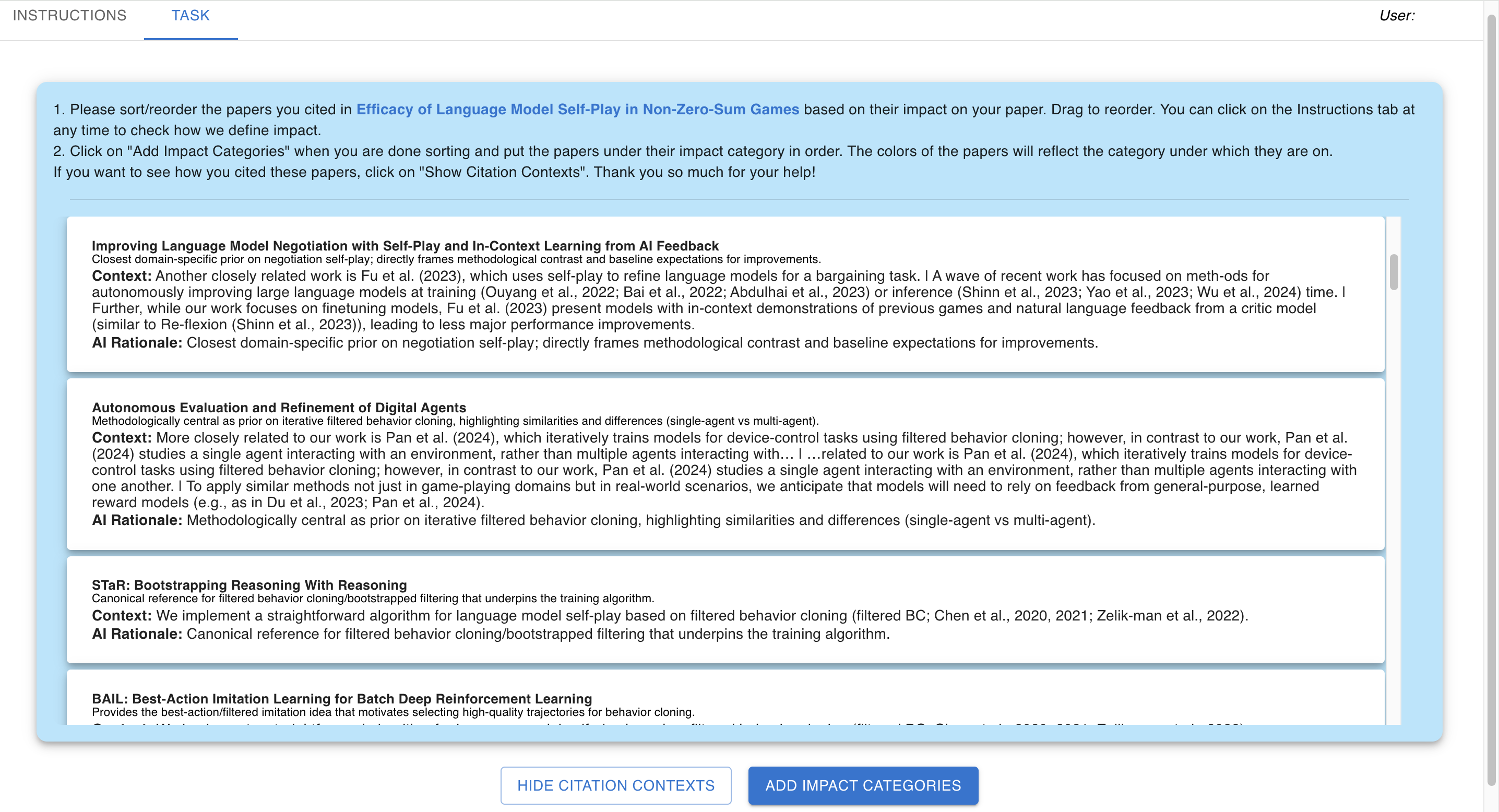}
  \caption{A feature of the interface that reveals the citation context when the "Show Citation Contexts" button is clicked, illustrating how each reference was cited in the paper.}
  \label{fig:Pilot_3}
\end{figure*}

\begin{figure*}[t]
  \centering
  \includegraphics[width=\textwidth]{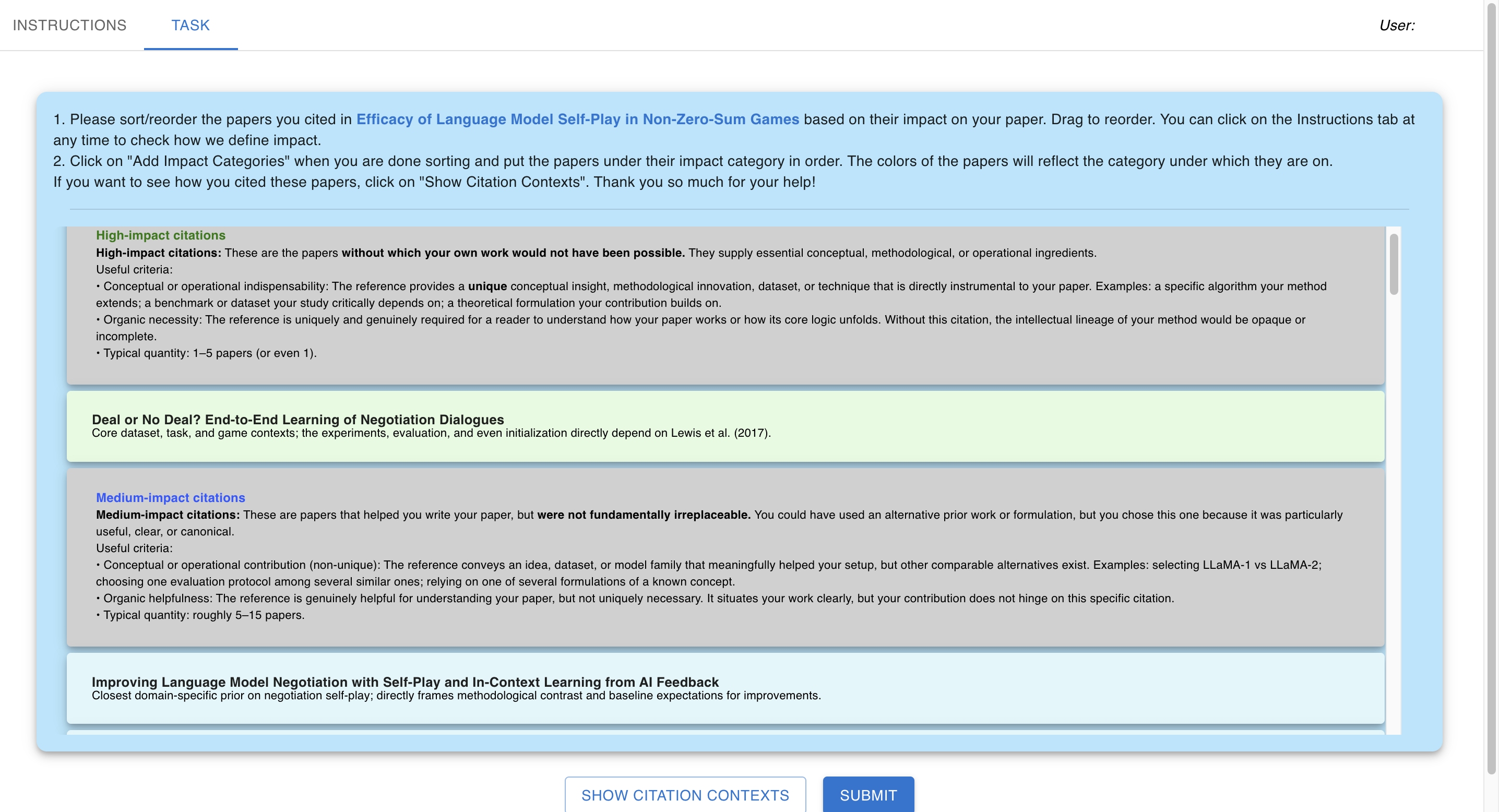}
  \caption{The final ranked list of references generated within the interface, showing the outcome of the annotation task with color-coded impact categories. Annotators submit their final ranked list of references.}
  \label{fig:Pilot_4}
\end{figure*}

Figures~\ref{fig:Pilot_1}--\ref{fig:Pilot_4} illustrate the custom annotation interface we designed for the study.  
Figure~\ref{fig:Pilot_1} shows the main annotation task layout.  
Figure~\ref{fig:Pilot_2} displays the complete list of references for an annotator's paper.  
Figure~\ref{fig:Pilot_3} demonstrates the feature we added to reveal cited papers' citation context. Lastly, 
Figure~\ref{fig:Pilot_4} presents the final ranked list generated within the interface.  As shown in Figure~\ref{fig:Pilot_4}, the impact definition for each category was available to the annotators throughout the task, and citations were color-coded according to their corresponding impact category.

\begin{figure*}[t]
  \centering
  \includegraphics[width=\textwidth]{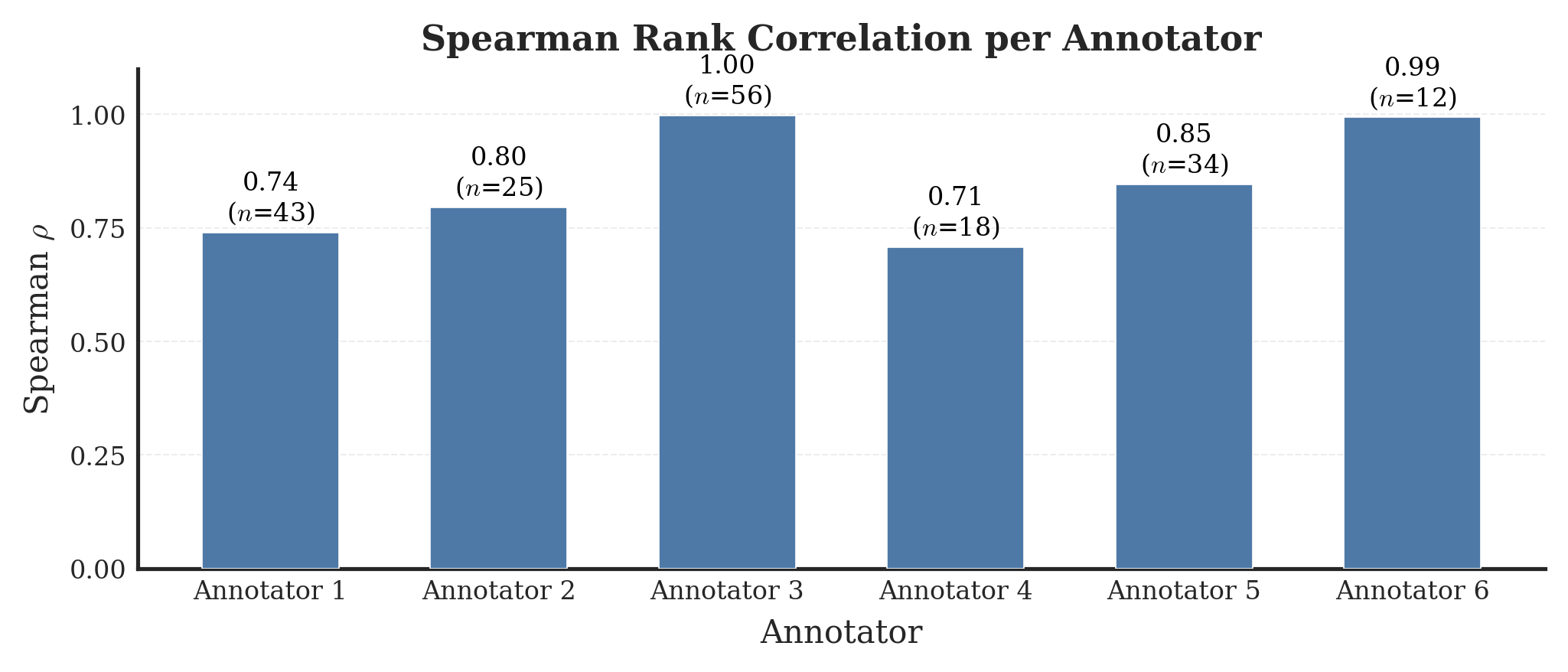}
  \caption{Shows the Spearman correlation between each annotator's ranking and the ranking generated by \textsc{Gpt}-5.1 using the same prompt from our experiments. The $n$ value above each bar represents the total number of references ranked in that paper. All correlation values exceeded $0.7$, indicating strong agreement.}
  \label{fig:PilotStudyResults}
\end{figure*}

We compute the Spearman rank correlation between each annotator's ranking and the ranking generated by \textsc{Gpt}-5.1 using the same prompt from our experiments (Figure~\ref{fig:ranking_prompt}). As shown in Figure~\ref{fig:PilotStudyResults}, all correlation values exceeded $0.7$, indicating strong agreement. These results suggest that our prompt produces rankings that closely align with how authors would rank their citations, though we defer a more comprehensive evaluation to future work.

\section{ACL Case Study Details}
\label{app:acl}

\begin{figure*}[t]
  \centering
  \includegraphics[width=\textwidth]{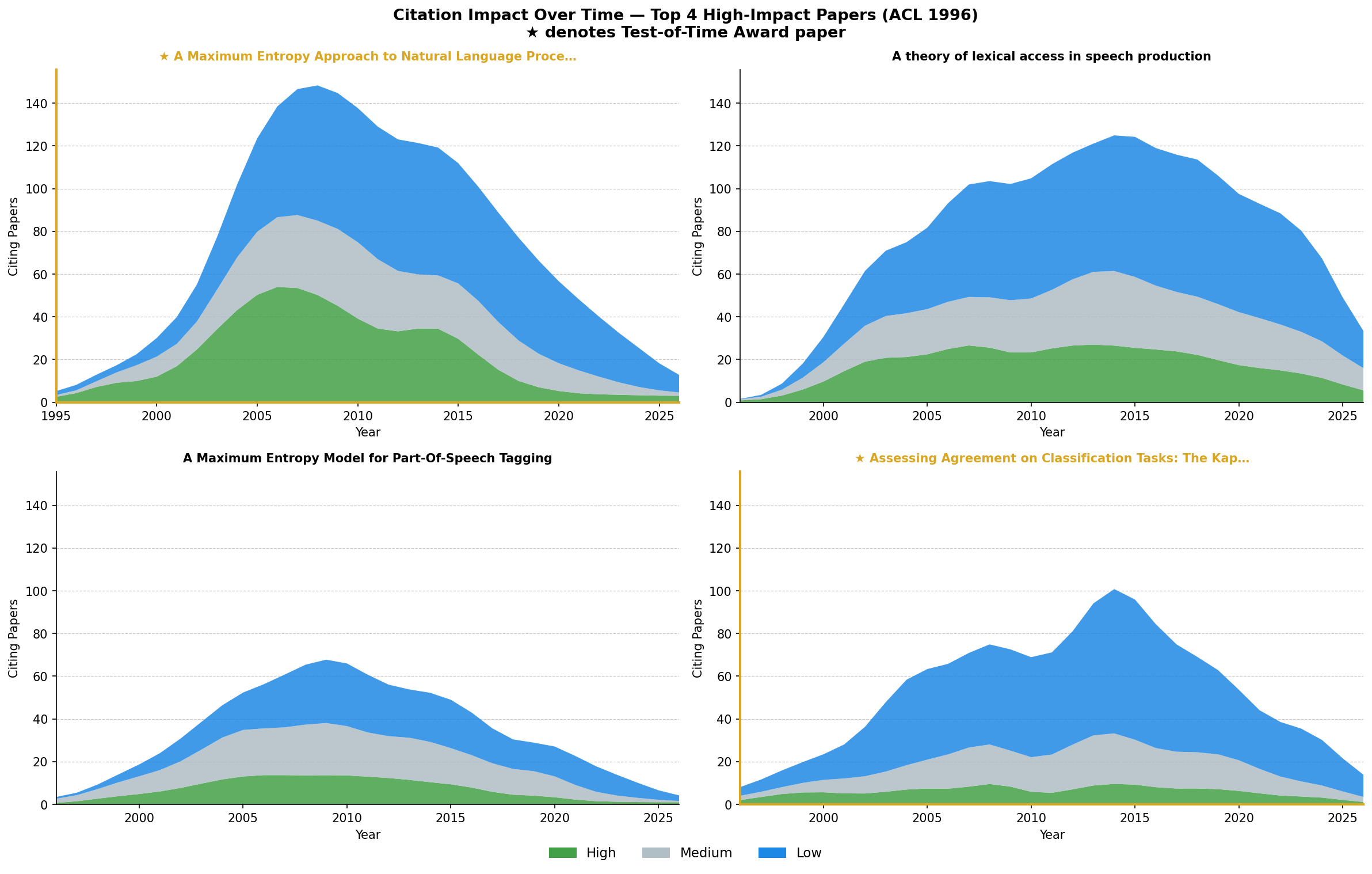}
  \caption{Yearly citation impact distribution for the top four ACL 1996 papers ranked by high-impact citations under \methodname{}. Each subplot stacks \textcolor[HTML]{43A047}{high}-, \textcolor[HTML]{546E7A}{medium}-, and \textcolor[HTML]{1E88E5}{low}-impact citations by year of the citing paper. Test-of-Time award winners (\citet{berger1996maximum}, top-left; \citet{carletta1996assessing}, bottom-right) are marked with a $\star$ and highlighted with a yellow border. \citet{berger1996maximum} accumulates the largest volume of high-impact citations of the four papers, sustained across more than two decades. \citet{carletta1996assessing} exhibits a stable but smaller stream of high-impact citations, with most of its citation volume concentrated in the low-impact category.}
  \label{fig:temporal-1996}
\end{figure*}

\begin{figure*}[t]
  \centering
  \includegraphics[width=\textwidth]{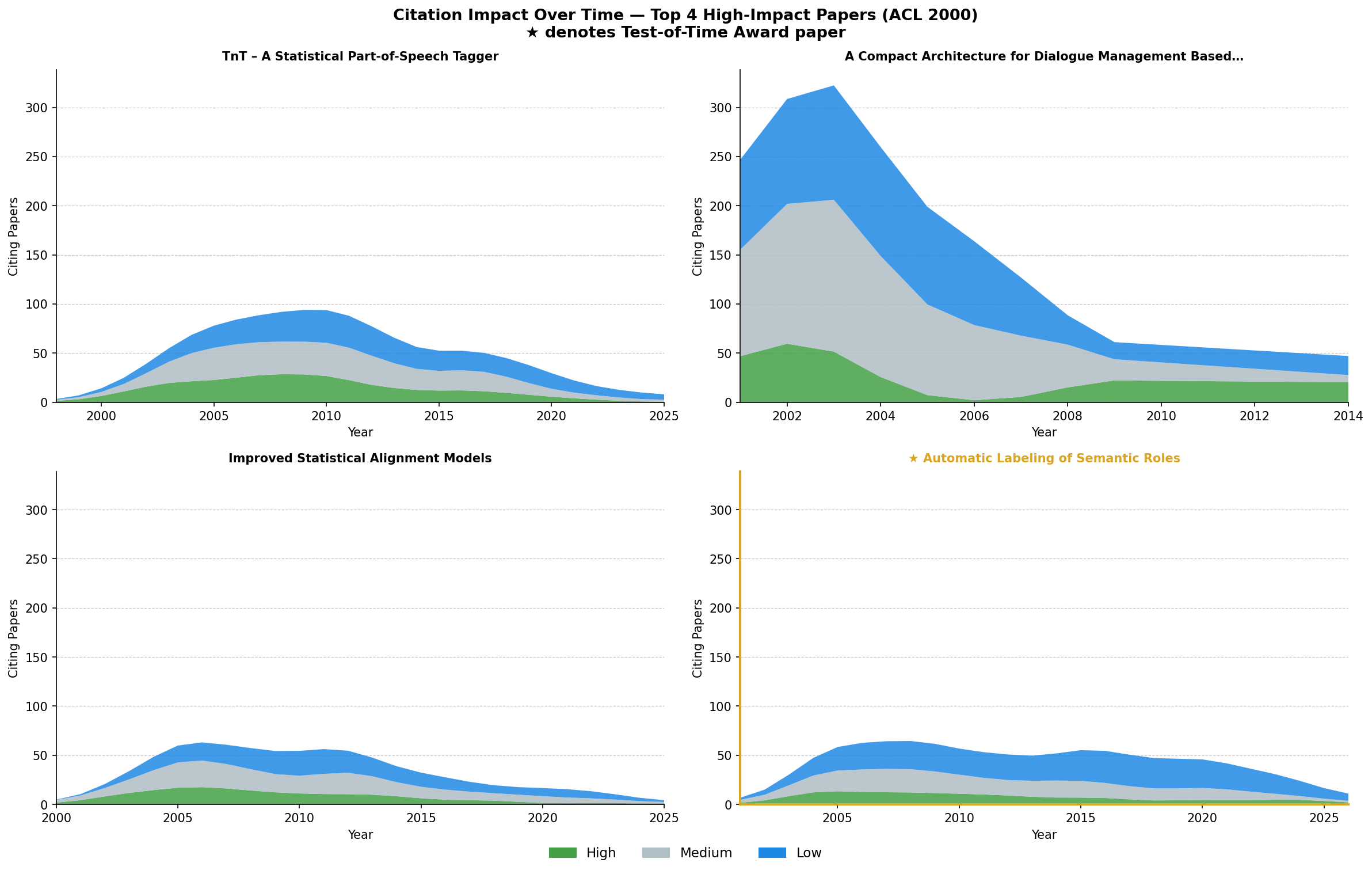}
  \caption{Yearly citation impact distribution for the top four ACL 2000 papers ranked by high-impact citations under \methodname{} (as shown in Figure~\ref{fig:ACL_results}). Each subplot stacks \textcolor[HTML]{43A047}{high}-, \textcolor[HTML]{546E7A}{medium}-, and \textcolor[HTML]{1E88E5}{low}-impact citations by year of the citing paper. The Test-of-Time award winner (\citet{gildea2002automatic}, bottom-right) is marked with a $\star$ and highlighted with a yellow border. Its high-impact citation profile remains stable through 2025, while several non-awarded top-ranked papers peak earlier and decline more steeply.}
  \label{fig:temporal-2000}
\end{figure*}

Table~\ref{tab:impact-distribution} reports the number of high-, medium-, and low-impact citations identified by \methodname{} in the ACL case study. Figures~\ref{fig:temporal-1996} and \ref{fig:temporal-2000} show the temporal distribution of these citation types for the top four papers in ACL 1996 and ACL 2000, respectively, ranked by their high-impact citation counts under \methodname{}.

\begin{table}[h]
\centering
\footnotesize
\begin{tabular*}{\columnwidth}{@{\extracolsep{\fill}}lrrr}
\toprule
\textbf{Dataset} & High & Medium & Low \\
\midrule
ACL 1996 & 3,934 (19\%) & 6,190 (30\%) & 10,696 (51\%) \\
ACL 2000 & 5,987 (17\%) & 12,184 (34\%) & 17,901 (50\%) \\
\bottomrule
\end{tabular*}
\caption{Distribution of impact labels assigned by \methodname{} across all citations of ACL 1996 and ACL 2000 publications. ``High'', ``Medium'', and ``Low'' report the number (and percentage) of citations assigned to each impact category.}
\label{tab:impact-distribution}
\end{table}

\section{Details about Models Used in Experiments}
As described in \S\ref{Sec:Experiments}, we use \textsc{Gpt-5.1} and o4-mini as our closed-source models, and Qwen3-30B-A3B-Instruct-2507-FP8 \cite{qwen3technicalreport} as our open-weight model. We adopt the recommended temperature and top-$p$ settings for each model (0.7 and 0.8 for Qwen; defaults for \textsc{Gpt-5.1} and o4-mini). These models all support large context windows of at least $200K$ tokens. We use these parameters for all experiments in \S\ref{Sec:Experiments}, \S\ref{app:two-labels}, and \S\ref{sec:tot-case-study}.

The total cost of running the experiments described in \S\ref{Sec:Experiments} was \$77.51 for the closed-source models, including approximately \$48 for the UKP experiments and \$30 for \methodname{} experiments. All experiments involving Qwen3-30B (in \S\ref{Sec:Experiments}, \S\ref{app:two-labels}, and \S\ref{sec:tot-case-study}) were conducted by hosting the model with vLLM on H100 GPUs. Although precise runtime measurements for all Qwen experiments are unavailable, based on our setup (H100 GPU with vLLM and FP8 inference), we estimate that the full set of experiments, i.e. running \methodname{} on \datasetname{} (46,890 papers), can be completed within approximately 10 days of continuous runtime on a single instance. This runtime could be further reduced by using multiple GPUs.

\section{\datasetname{}}
\label{crystal-bank}
\datasetname{} contains \datasetsize{} citing papers with rankings and impact labels produced by \methodname{}. These correspond to the citing papers used in \S\ref{Sec:Experiments} and \S\ref{sec:tot-case-study}, comprising 442 and 46{,}448 papers, respectively.

\section{Disclosure on the Use of Generative Assistants}
\label{sec:ai_assistance}
The authors adhered to ACL's guidelines for appropriate use of generative assistance in authorship. In particular, we used generative assistants to polish our original writing. Additionally, we used LLM-powered tools, such as \cite{asta2025}, for literature search.

\end{document}